\documentclass[twocolumn]{aastex63}
\usepackage{natbib}
\usepackage{tabularx}

\def\subinrm#1{\sb{\mathrm{#1}}}
{\catcode`\_=13 \global\let_=\subinrm}
\mathcode`_="8000
\def\supinrm#1{\sp{\mathrm{#1}}}
{\catcode`\^=13 \global\let^=\supinrm}
\mathcode`^="8000
\def\upsubscripts{\catcode`\_=12 } 
 
\begin{document}
\upsubscripts

\title{PLANETESIMAL ACCRETION AT SHORT ORBITAL PERIODS}

\author{Spencer C. Wallace}
\affiliation{Astronomy Department, University of Washington, Seattle, WA 98195}

\author{Thomas R. Quinn}
\affiliation{Astronomy Department, University of Washington, Seattle, WA 98195}

\begin{abstract}
Formation models in which terrestrial bodies grow via the pairwise accretion of planetesimals have been reasonably successful at reproducing the general properties of the solar system, including small body populations. However, planetesimal accretion has not yet been fully explored in the context of the wide variety of recently discovered extrasolar planetary systems, particularly those that host short-period terrestrial planets. In this work, we use direct N-body simulations to explore and understand the growth of planetary embryos from planetesimals in disks extending down to $\simeq$ 1 day orbital periods. We show that planetesimal accretion becomes nearly 100 percent efficient at short orbital periods, leading to embryo masses that are much larger than the classical isolation mass. For rocky bodies, the physical size of the object begins to occupy a significant fraction of its Hill sphere towards the inner edge of the disk. In this regime, most close encounters result in collisions, rather than scattering, and the system does not develop a bimodal population of dynamically hot planetesimals and dynamically cold oligarchs, like is seen in previous studies. The highly efficient accretion seen at short orbital periods implies that systems of tightly-packed inner planets should be almost completely devoid of any residual small bodies. We demonstrate the robustness of our results to assumptions about the initial disk model, and also investigate the effects that our simplified collision model has on the emergence of this non-oligarchic growth mode in a planet forming disk.
\end{abstract}

\section{Introduction} \label{sec:intro}

Planetesimal accretion is a key phase in the terrestrial planet growth
process, bridging the gap from kilometer-sized bodies up to roughly
moon-sized objects known as planetary embryos. In the earliest stages
of the planet formation process, beginning from $\mu$m sizes, aerodynamic forces dominate the
growth and evolution of the solids and statistical models
\citep{johansen14, birnstiel16} are appropriate to describe how these
numerous, small bodies coagulate. 

Due to the internal pressure support
of the gas disk, the gas itself orbits at sub-Keplerian speed and
exerts a headwind on any solids large enough to decouple from the gas
\citep{weidenschilling77}. Around a meter in size, this headwind
is maximally effective at sapping away orbital angular momentum, and planet-building material can fall onto the central star on 
catastrophically short timescales \citep{weidenschilling77, nakagawa86}. Additionally, laboratory experiments suggest that 
collisions between mm- to cm- sized solids tend to result in bounces or destruction, rather than continued growth
\citep{blum93, colwell03, beitz11}.

For these reasons, a number of mechanisms to radially concentrate solids in a planet-forming disk have been proposed to 
facilitate fast growth from mm to km sizes \citep{johansen07, lyra08, bai10} in order to surmount these barriers. Interestingly, 
formation models for the short-period multiplanet systems revealed by Kepler \citep{fabrycky14} also seem to require enhanced 
concentrations of planet-building material to reproduce the observed architectures \citep{raymond07, hansen12}.

Regardless of how the mm- to km-sized growth barriers are surmounted, gravity begins to dominate and aerodynamic gas drag 
plays an increasingly unimportant role beyond this size. During this phase, collision cross sections are enhanced as gravitational 
focusing \citep{safronov69} acts to bend the trajectories of bodies undergoing close encounters. Because gravitational focusing 
becomes more effective as bodies grow larger, a period of runaway growth occurs \citep{wetherill89, kokubo96, barnes09} and a 
power law spectrum of masses develop. Eventually, the largest bodies (known as oligarchs) dynamically heat the remaining 
planetesimals, severely limiting further growth \citep{kokubo98}. The final outcome of this phase is a bimodal population of 
dynamically cold oligarchs, surrounded by dynamically hot, difficult to accrete residual planetesimals. Lines of evidence suggest 
that the asteroid belt \citep{bottke05, morbidelli09}, Kuiper belt \citep{duncan89, levison08, sheppard10} and the Oort cloud  \citep{levison11} are largely composed of the leftovers of this stage of planet formation.

Tidal interactions between protoplanets and the gaseous disk keep eccentricities and inclinations low until the gas disk dissipates, typically over the course of a few Myr \citep{mamajek09}. On a timescale of roughly $10^{5}$ yr, gravitational perturbations trigger an instability which involves large scale oscillations of the eccentricity and inclinations of bodies as they strongly interact through secular resonances \citep{chambers98}.
As a consequence of the instability, the oligarchics are no longer
on isolated, stable orbits and coalesce to form Earth-sized planets
through a series of extremely energetic giant 
impacts \citep{kokubo02, raymond05, raymond06}. 

Due to the relative ease of modeling the early dust coagulation phases
and the final giant impact phase, these steps in the terrestrial
planet formation process have received most of the attention in the
literature. The planetesimal accretion phase, which we will focus on
in this paper, is more difficult and expensive to model because there are too
many particles to directly track with traditional N-body codes, while
the gravitational interactions between the few massive bodies produced by the runaway growth phase \citep{ida93, kokubo95, kokubo98} render statistical methods inappropriate. Due to computational expense, N-body simulations of
planetesimal accretion are usually modeled in a narrow
ring \citep{kokubo96, kokubo98}, and the results and timescales are then scaled to suit whatever relevant orbital period is being 
studied. N-body simulations of terrestrial planet assembly typically begin with a series of neatly spaced oligarchs, whose mass 
varies smoothly with orbital period. As we will show in this paper, oligarchic growth does not scale to arbitrarily short
orbital periods.

Given that Systems of Tightly-packed Inner Planets (STIPs) appear to
be a common outcome of planet formation \citep{lantham11, lissauer11, rowe14}, understanding exactly how
solids accumulate at short orbital periods is crucial. Although
gas-disk driven migration of the planets themselves is often invoked
to explain the observed architectures \citep{izidoro17, izidoro21}, we will focus on an in-situ
model in this paper. That is, once the planetesimals themselves form,
they largely stay in place, and any subsequent large-scale movement of
the solids are the result of mutual gravitational interactions. The
focus of this work will be to understand how the outcome of the
planetesimal accretion process scales with orbital period by using a
high-powered N-body code to directly follow the growth and evolution
of the planetesimals across a wide range of orbital periods (1 to 100
days). In doing so, we will assess whether the typical initial conditions (fully formed, evenly spaced protoplanets, e.g. \citet{raymond06}) used in studies of terrestrial planet formation are actually appropriate for understanding STIPs.

In section \ref{sec:theory} we provide an overview of the theory
behind planetesimal accretion and show that assumptions used to derive
the well-known modes of growth are only valid at sufficiently long
orbital periods. We then motivate the need for N-body simulations to
study this problem and describe the code used, along with how our
initial conditions were constructed in section \ref{sec:methods}. In
section \ref{sec:narrow}, we present a parameter study of planetesimal
accretion using a series of simulations of narrow annuli that exhibit both oligarchic and non-oligarchic growth. In section 
\ref{sec:fulldisk} we present a set of simulations starting with a  wide planetesimal disk and demonstrate that a transition 
between accretion modes occurs at moderately small orbital periods. Next, we assess the impact of 
simplifications made to our collision model on this result in section \ref{sec:assump}. In section \ref{sec:discuss}, we discuss the 
implications of this multimodal accretion behavior throughout the disk for planet formation models and conclude.

\section{Overview of Planetesimal Accretion}\label{sec:theory}

\subsection{Oligarchic and Runaway Growth}

We begin our analysis by considering a disk of equal mass planetesimals
with radius $r_{pl}$, mass $m_{pl}$ and surface density
$\Sigma_{pl}$. The collision rate in the vicinity of an orbit defined
by Keplerian frequency $\Omega$ can be written as $n \Gamma v$, where
$n = \Sigma_{pl} \Omega / 2 m_{pl} v$ (where we have assumed that the scale height of the planetesimal disk goes as $2v/\Omega$). $\Gamma$ describes the effective
collision cross section and $v$ is the typical encounter velocity
between planetesimals.
For a swarm of planetesimals on randomly oriented orbits, $v$ is typically
taken to be the rms velocity, $\left< v^{2} \right>^{1/2}$, which can be related to the eccentricity and inclination distribution $(e, i)$ in the following way \citep{lissauer93}:
\begin{equation}\label{eq:ecc_vel}
	\left< v^{2} \right>^{1/2} = \left( \frac{5}{4} \left< e^{2} \right>^{1/2} + \left< i^{2} \right>^{1/2}  \right) v_{k}.
\end{equation}

The dynamical interactions between growing planetesimals can be somewhat simplified by scaling the orbital elements of the bodies by
the Hill radius

\begin{equation}\label{eq:hillfac}
	r_{h} = \left(\frac{m_{pl}}{3 M_{*}}\right)^{1/3}, 
\end{equation}
\noindent where $M_{*}$ is the mass of the central star. The Hill radius of a body describes the size scale over which the gravity of the growing planetesimal dominates over the gravity of the star. Using equation \ref{eq:hillfac}, the eccentricity, inclination and separation between orbiting bodies can be defined as

\begin{equation}\label{eq:hillorb}
	e_{h} = \frac{a e}{r_{h}}, \: i_{h} = \frac{a i}{r_{h}}, \: \tilde{b} = \frac{a_{2} - a_{1}}{r_{h}}.
\end{equation}

\noindent Using this formalism, $e_{h}$ and $i_{h}$ describe the radial and vertical excursions of an orbiting body in units of its Hill radius. For $e_{h} > 1$, the random velocity dispersion dominates over the shearing motion across a separation of $1 r_{h}$ and encounters can be treated with a two-body formalism.

Assuming that every collision results in a perfect merger, the growth rate of a planetesimal is given by
\begin{equation}\label{eq:growth}
	\frac{1}{M}\frac{dM}{dt} = \frac{\Sigma \Omega}{2 m_{pl}} \Gamma.
\end{equation}
\noindent In the case where the collision cross section depends only
on the physical size of the planetesimals, the growth scales sub-linearly
with mass and the mass distribution is expected to evolve in an
``orderly'' fashion, in which mass ratios between bodies tend toward unity. However, bodies larger than $\sim 100$ km in size are expected to exert a significant gravitational force on each other during encounters and the collision cross section depends on both the size of the bodies and their encounter velocities. In this case, 
\begin{equation}\label{eq:gravfocus}
	\Gamma = \pi r_{pl}^2 \left( 1 + v_{esc}^2 / v^2 \right)
\end{equation}
\citep{safronov69}, where $v_{esc}$ is the escape velocity from the two bodies at the point of contact.

In the limit that $v_{esc} \gg v$, it can be shown that $dM/dt \propto
M^{4/3}$, which implies a runaway scenario where growth
accelerates with mass. This mode of growth was confirmed with N-body
simulations by \citet{kokubo96} and appears necessary to construct
protoplanets within the lifetime of a protoplanetary disk \citep{lissauer87}, although one should note that pebble accretion \citep{lambrechts12, lambrechts14, bitsch15} is a viable alternative scenario. Due to the
velocity dependence of the gravitational focusing effect, it is not clear how ubiquitous this mode of growth is. In particular, encounter velocities at short orbital periods will be rather large (because $v \sim v_{k}$) and the $v_{esc} \gg v$ condition may not always be satisfied. The effect that a dynamically hot disk has on runway growth will be examined in detail in section \ref{sec:narrow}.

An important feature is missing from the model described above, which
limits its applicability at late times. Gravitational stirring, which converts
Keplerian shear into random motion, raises the typical encounter velocity
between planetesimals over time \citep{weidenschilling89, ida90} and diminishes the effectiveness of gravitational
focusing. As the mass spectrum of the system evolves away from uniformity, these velocity differences become even more
pronounced. As the system evolves, it tends toward a state of
energy equipartition where $v \sim m^{1/2}$. For a system of equal mass bodies in which encounters are driven by random 
motions rather than Keplerian shear (dispersion dominated), the timescale for gravitational stirring is described by the two-body 
relaxation time \citep{ida93}
\begin{equation}\label{eq:relax}
	t_{relax} = \frac{v^3}{4 \pi n G^2 {m_{pl}}^2 \ln \Lambda},
\end{equation}
where $\ln \Lambda$ is the Coulomb logarithm,
typically taken to be $\approx 3$ for a planetesimal disk \citep{ida90, stewart00}. Despite
the fact that the behavior of gravitational stirring is well-described
by a two-body formula, \citep{ida93} found that the stirring in a planetesimal disk is actually driven by close encounters, which 
requires a three-body formalism. As we will show in section \ref{sec:narrow}, gravitational stirring effectively shuts off when the 
Hill sphere of a body becomes comparable to its physical size. In this case, close encounters tend to result in collisions, and the 
main pathway for energy exchange between planetesimals and growing protoplanets is unable to operate.

\citet{kokubo98} showed that the runaway growth process described above is actually self-limiting. As the runaway
bodies develop, they become increasingly effective at dynamically heating the
remaining planetesimals, which diminishes the
gravitational focusing cross sections and throttles the growth rate. Around the time
that the mass of the runaway bodies exceeds the mass of the planetesimals
by a factor of $\sim 50-100$
\citep{ida93} a phase of less vigorous ``oligarchic'' 
growth commences, in which the largest bodies continue to 
accrete planetesimals at similar rates, independent of mass.

Regardless of the mechanism that eventually limits the growth of the planetary embryos, a maximum estimate for the masses produced during this 
stage of accretion can be obtained using the initial solid surface density profile. A growing protoplanet is expected to accrete material within a distance
$b = \tilde{b} r_{h}$ of its orbit. The total mass of planetesimals
within this distance is then $2 \pi a \left(2 \tilde{b} r_{h} \right)
\Sigma_{pl}$, where $a$ is the semimajor axis of the growing protoplanet. The ``isolation mass'' of the protoplanet can then be obtained by setting the protoplanet mass equal to the total mass of planetesimals within accretionary reach such that

\begin{equation}\label{eq:iso_mass1}
	M_{iso} = 4 \pi a^{2} \tilde{b} \left(\frac{M_{iso}}{3 M_{*}} \right)^{1/3} \Sigma_{pl}.
\end{equation}
\noindent Solving for $M_{iso}$ gives
\begin{equation}\label{eq:iso}
	M_{iso} = \left[ \frac{\left( 2 \pi a^2 \Sigma_{pl} \tilde{b} \right)^3}{3 M_{*}} \right]^{1/2}.
\end{equation}
\noindent For bodies on circular, non-inclined orbits, $\tilde{b} = 2 \sqrt{3}$ is the smallest orbital separation that produces a non-negative
Jacobi energy and permits a close encounter \citep{naka88}. This value of $\tilde{b}$ is typically used to calculate the final isolation mass of a protoplanet because oligarchic growth tends to maintain near-circular orbits for the growing protoplanets.

The picture described above relies upon a crucial assumption, which
is that the mass distribution evolves slowly enough for gravitational stirring to maintain energy equipartition. In other words, the relaxation timescale must remain short relative to the collision timescale. For typical conditions near the terrestrial region of the solar system, this timescale condition is satisfied. Due to the steep dependence of the relaxation time on encounter velocity, however, this condition can easily be violated at shorter orbital periods.

\begin{figure}
\begin{center}
    \includegraphics[width=0.5\textwidth]{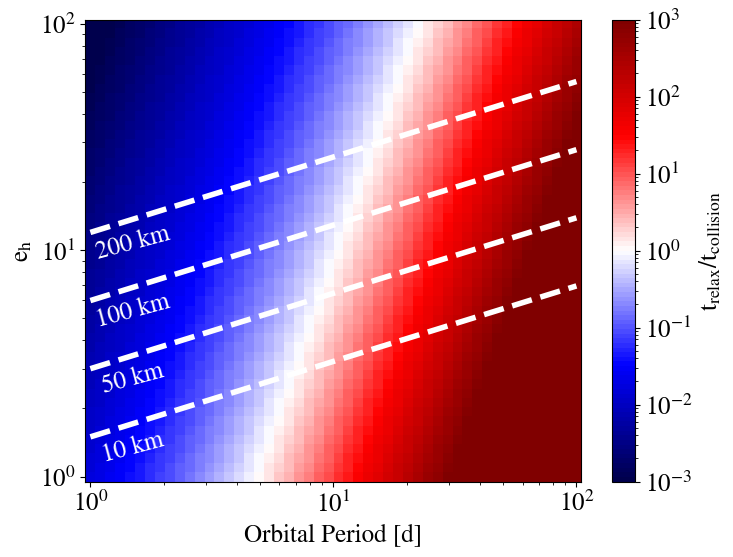}
    \caption{The ratio between the two-body relaxation and collision
      timescale for a population of equal-mass planetesimals with an
      internal density of 3 g cm$^{-3}$. The dashed curves show 
      the value of $e_{h}$ for which the random velocity dispersion is equal
      to the escape velocity of the planetesimals for a range of sizes. Only
      in regions where $t_{relax} \gg t_{coll}$ can the velocity distribution 
      respond to changes in the mass of the bodies such that oligarchic 
      growth can operate. This condition is no longer satisfied for a 
      dynamically hot disk at sufficiently short orbital periods.\label{fig:timescales}}
\end{center}
\end{figure}

In figure \ref{fig:timescales}, we show the ratio between the relaxation
and collision timescale for a population of equal-mass planetesimals
as a function of orbital period. Here, the encounter velocity is
described by equation \ref{eq:ecc_vel}. For
simplicity, we assume that $\left< e^2 \right>^{1/2} = 2\left< i^2
\right>^{1/2}$ \citep{ida93a} and that the eccentricity dispersion is
constant with orbital period. The coloring indicates the ratio between
$t_{relax}$ and $t_{coll}$. The dashed lines denote the eccentricity at which the
random encounter velocity (calculated according to equation \ref{eq:ecc_vel}) is
equal to the mutual escape velocity of the bodies. This is shown for planetesimals
with an internal density of 3 g cm$^{-3}$ ranging from 10 to 200 km in size.

A physically realistic value of $e_{h}$ for a population of planetesimals is going to depend on the structure
of the gaseous disk (which we will address in section \ref{sec:fulldisk}). However, the eccentricity
dispersion will typically increase over time until $\left< v^{2} \right>^{1/2} = v_{esc}$ and this is
often used to construct the initial conditions (e.g. \citet{barnes09}). Therefore, the curves in this figure
should be interpreted as upper limits.

The timescale criterion for oligarchic growth is only satisfied in regions where the
disk is sufficiently dynamically cold and the orbital period is
sufficiently long. In sections \ref{sec:narrow} and \ref{sec:fulldisk}
we will explore the behavior and outcome of planetesimal accretion in regions where this criterion is \textit{not} satisfied. 

\subsection{Planetesimal Size and Extent of Hill Sphere}\label{sec:sizeandhill}

In the formalism described above, the mass and velocity distribution
of the bodies are both a function of time. Due to the interdependence
of these quantities, it is not clear whether the ratio between the relaxation and collision timescales will remain constant as the
oligarchs develop. In the case of the solar system, $t_{relax} \ll t_{coll}$ likely continued to remain true, otherwise runaway 
growth would have consumed all of the small bodies and there would be nothing left to populate the asteroid or Kuiper belt. In 
the case where $t_{coll} \ll t_{relax}$, however, it is not clear how the system might evolve.

An insight into the expected behavior in this regime can be gained by
defining a dimensionless parameter, $\alpha$, which is the ratio
between the physical size of a body and its Hill radius, $r_{h}$
\begin{equation}\label{eq:alpha}
	\alpha = \frac{r_{pl}}{r_{h}} = \frac{1}{a} \left( \frac{9 M_{\star}}{4 \pi \rho_{pl}} \right)^{1/3},
\end{equation}
where $a$ is the semimajor axis of the body and $\rho_{pl}$ is its
bulk density. Assuming that the bulk density stays constant as bodies collide and
grow, and that no large-scale migration occurs, the scaling of both
$r_{pl}$ and $r_{h}$ as $m_{pl}^{1/3}$ means that $\alpha$ will be
constant with time. For a composition of ice and rock, $\alpha$ is
small for any presently populated region of the solar system ($\alpha \sim
10^{-2}$ near Earth and $\alpha \sim 10^{-4}$ in the Kuiper belt). As
one moves closer to the sun, $\alpha$ becomes larger than 1, which
implies that the Hill sphere of a body becomes smaller than its physical size.
As an additional note, the Roche limit of the central body and the distance at which $\alpha = 1$ are equivalent for a rigid spherical body. After applying a hydrostatic correction to the Roche limit, $a_{Roche}$ is equivalent to 0.6 times this distance. This accretion mode should therefore be relevant for planetary ring systems, which is a topic that deserves further study using high resolution N-body techniques.

The magnitude of $\alpha$ controls the relative importance of gravitational scattering and collisions in driving the evolution of the planetesimal disk. When $\alpha$ is small, most close encounters will result in a gravitational interaction, moving the system toward a state of relaxation. If, however, the Hill sphere is largely filled by the body itself, these same encounters will instead drive evolution of the mass spectrum. Because $\alpha$ stays constant with mass, the boundary in the disk where collisions or gravitational encounters dominate, will stay static with time.

We also introduce a second dimensionless quantity, which relates the physical size of the bodies to the velocity state of the system
\begin{equation}\label{eq:beta}
	\beta = \frac{r_{pl}}{r_{g}}.
\end{equation}
where $r_{g} = G m_{pl} / v^{2}$ is the gravitational radius of a body (see eq 4.1 of \citet{ida90}). Encounters between bodies inside of a distance of $r_{g}$ result in significant deflections of their trajectories. It should be noted that the gravitational focusing enhancement factor $v^{2}/v_{esc}^{2}$ is equal to 1 for $\beta = 1$. In the case where $r_{g}$ is smaller than the size of a planetesimal, the gravitational focusing enhancement factor will be between 0 and 1 and the collision cross section is mostly set by the geometric value. For very large values of $r_{g}$ ($\beta \ll 1$), the effective collision cross section is almost entirely set by gravitational scattering.

\begin{figure}
\begin{center}
    \includegraphics[width=0.5\textwidth]{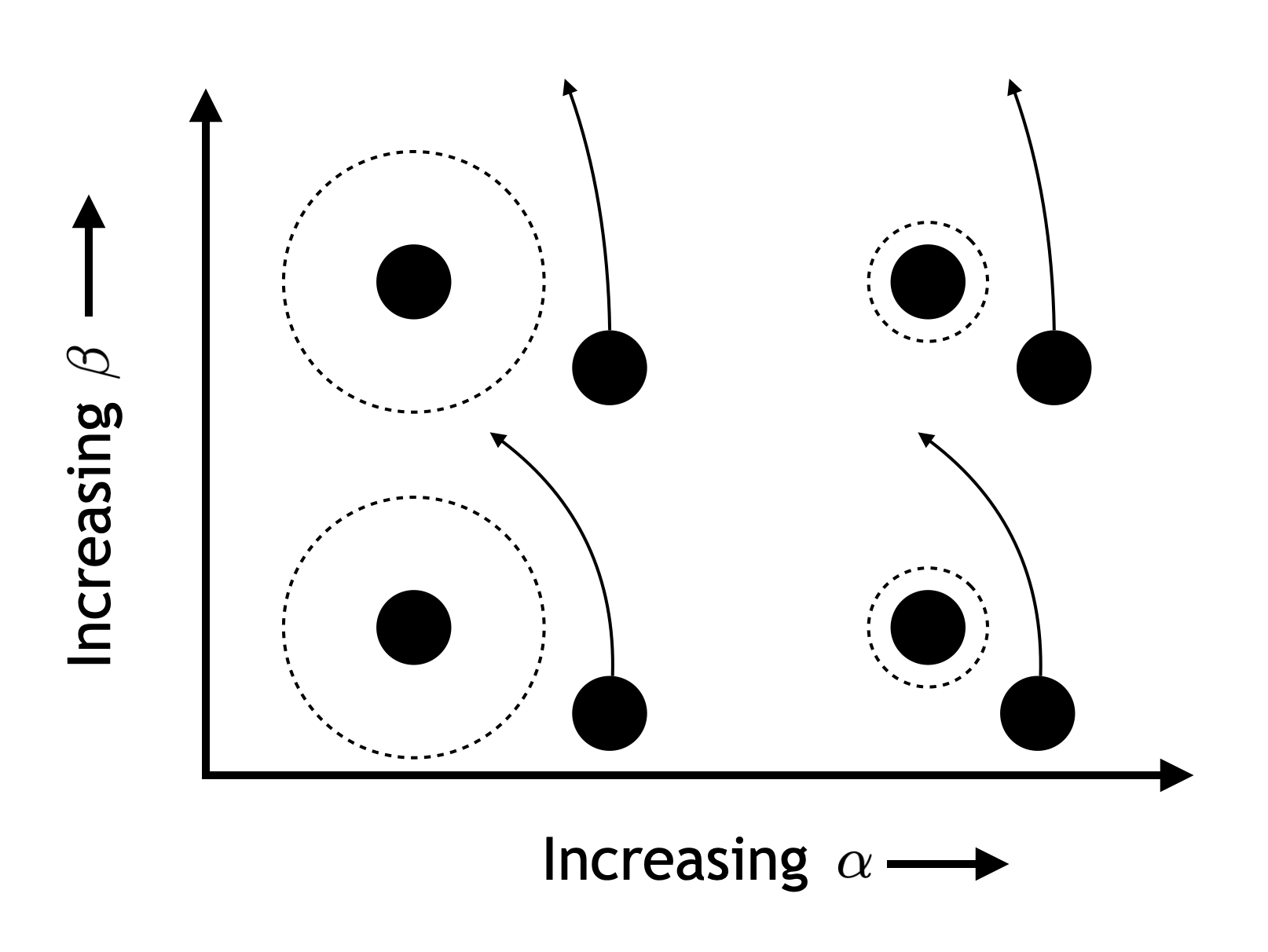}
    \caption{A diagram detailing how varying the values of $\alpha$ and $\beta$ affect the geometry of a two-body encounter. The solid circles represent the physical radius of a body and the dashed circles represent the Hill radius. As $\alpha$ is increased, the Hill radius and the physical radius become comparable in size. As $\beta$ is increased, the trajectory of a passing body for a fixed impact parameter is less affected.
      \label{fig:alpha_beta_diagram}}
\end{center}
\end{figure}

These scaling considerations motivate the range of parameters we choose for the numerical experiments presented in the next section, where we aim to understand where and when runaway and oligarchic growth can operate. In figure \ref{fig:alpha_beta_diagram}, we show a schematic which relates $\alpha$ and $\beta$ to the geometry of a two-body encounter. For large values of $\alpha$, the Hill radius of a body (dashed circle) becomes comparable to its physical radius (solid circle). As $\beta$ increases, the trajectory of a body undergoing a flyby is less affected by the encounter.

\begin{figure*}
\begin{center}
    \includegraphics[width=\textwidth]{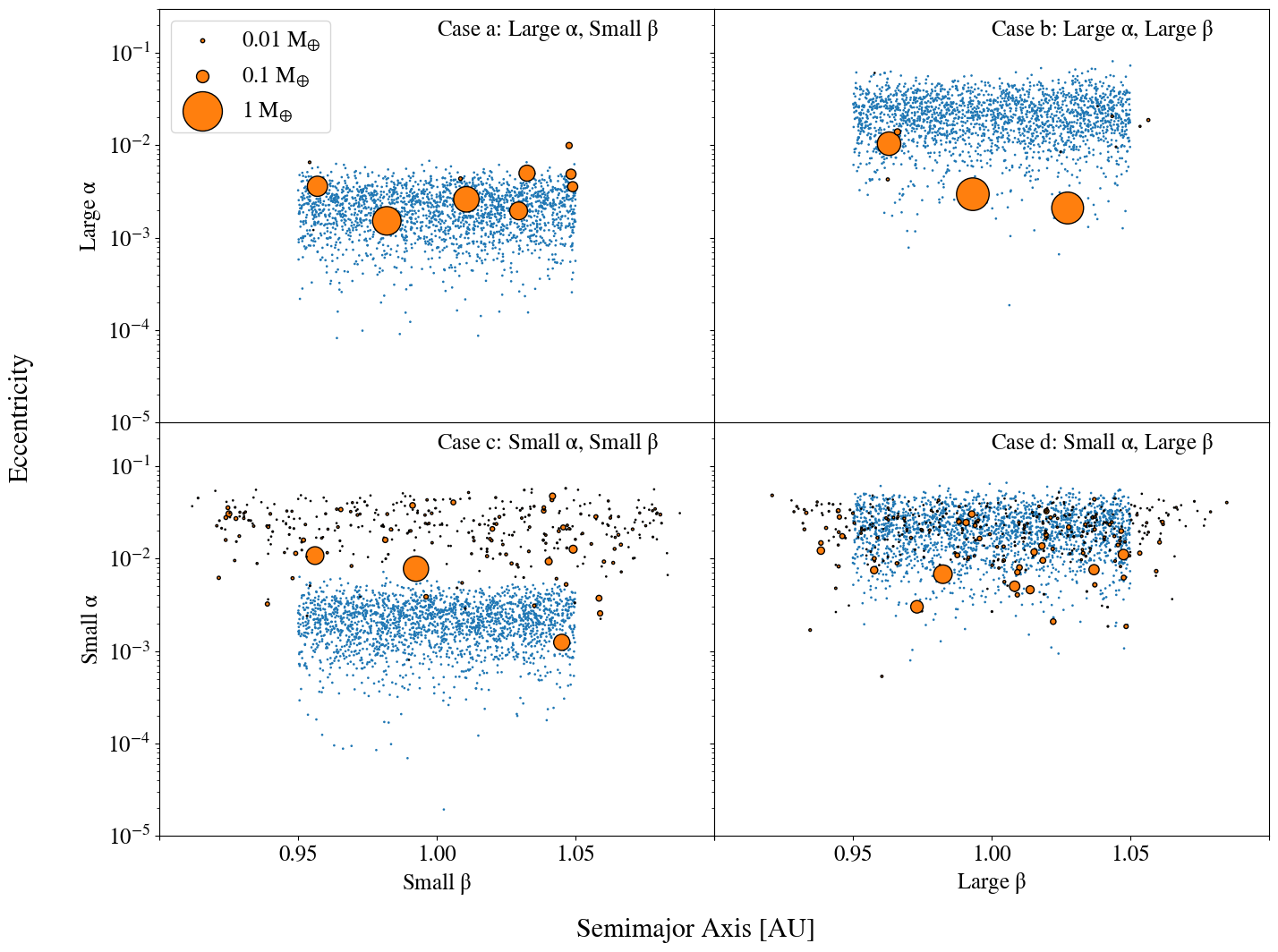}
    \caption{The initial (blue) and final (orange) states of the narrow annulus simulations described in section \ref{sec:narrow}. 
    Relative masses of the bodies are indicated by point size. In the case of large $\alpha$, almost no residual planetesimal 
    population remains. Regardless of the initial choice of $\beta$, the protoplanets that form attain similar eccentricities. 
    \label{fig:alpha_beta}}
\end{center}
\end{figure*}

\begin{figure}
\begin{center}
    \includegraphics[width=0.5\textwidth]{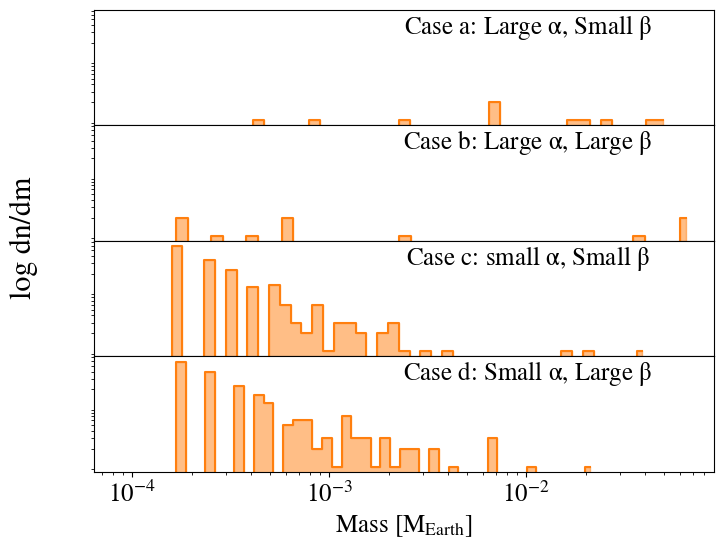}
    \caption{The final state of the mass distributions for the
      narrow annulus simulations described in section \ref{sec:narrow}. For small
      $\alpha$, a few embryos form alongside a power law tail of
      planetesimals. For larger values of $\alpha$, the mass distribution stops being bimodal. 
      As in the previous figure, the initial choice of $\beta$ does not appear to have any meaningful impact on the end result.
      \label{fig:alpha_beta_mass}}
\end{center}
\end{figure}

\begin{figure}
\begin{center}
    \includegraphics[width=0.5\textwidth]{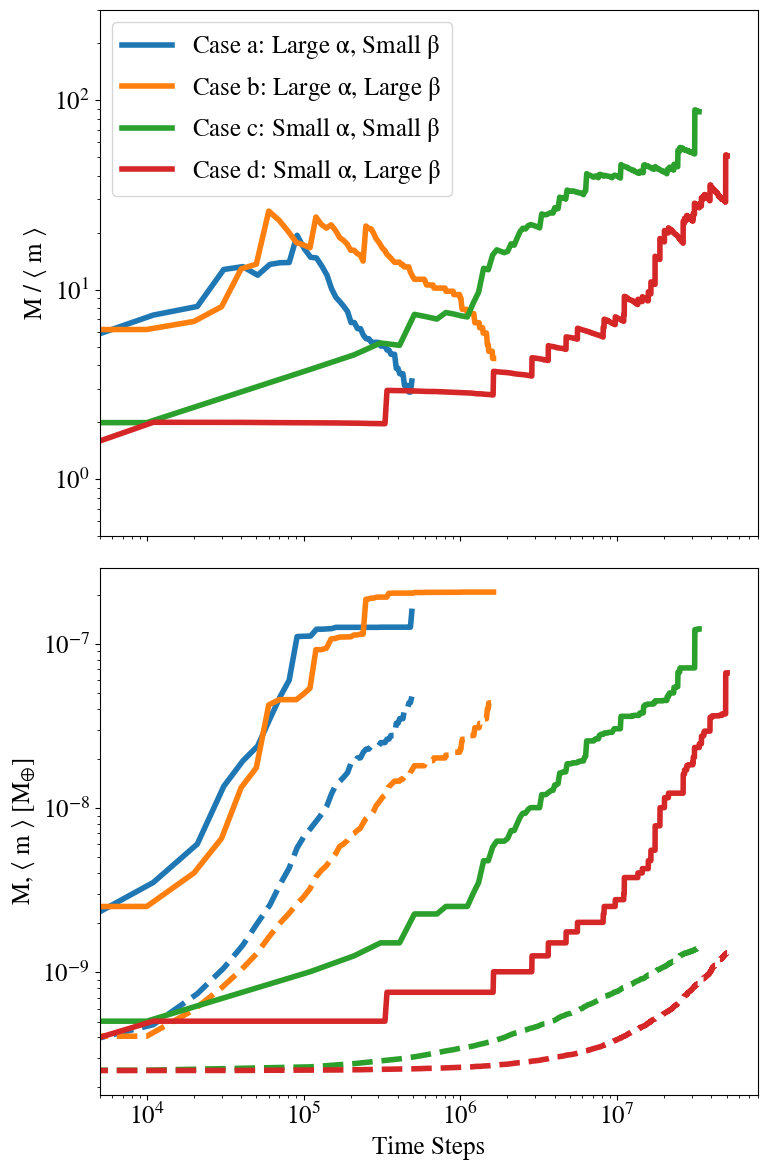}
    \caption{Top: The evolution of the ratio between the maximum and mean mass for the four simulations presented
    in section \ref{sec:narrow}. The runaway growth phase can be identified by a positive trend in this ratio. For all values of 
    $\alpha$, an increase in $\beta$ has the effect of delaying runaway growth. Bottom: The evolution of the maximum
    (solid lines) and mean (dashed lines) shown individually.\label{fig:alpha_beta_evo}}
\end{center}
\end{figure}

\section{Numerical Methods}\label{sec:methods}

We use the tree-based N-body code {\sc ChaNGa} to model the gravitational and collisional evolution of a swarm of planetesimals. {\sc ChaNGa} is written using the {\sc CHARM++} parallel programming language and has been shown to perform well on up to half a million processors \citep{menon15} and can follow the evolution of gravitationally interacting collections of up to billions of particles. Using a modified Barnes-Hut tree with hexadecapole expansions of the moments to approximate forces, {\sc ChaNGa} integrates the equations of motion using a kick-drift-kick leapfrog scheme. For all of the simulations presented in this paper, we use a node opening criteria of $\theta_{BH}$ = 0.7. Additional information about the code is available in \citep{jetley08,menon15}.

Using the neighbor-finding algorithm in {\sc ChaNGa}, originally
designed for SPH calculations, we have recently implemented a solid
body collision module in the code.  This work is largely based on the
solid-body collision implementation in {\sc PKDGRAV}, which is
described in \citet{richardson94} and \citet{richardson00}. To
summarize, imminent collisions are detected during the ``drift'' phase
by extrapolating positions of bodies forward in time, using the
velocity calculated at the opening ``kick''. For each body, any
neighboring particles which fall within a search ball of radius
$2 \Delta T v + 2 r_{pl}$, where $\Delta T$ is the current timestep size
for the particle and $v$ is magnitude of its heliocentric velocity,
 are tested for an imminent collision. In the case that
a collision is detected, the particles are merged into a single larger
body, which is given the center of mass position and velocity of the
two parents. Resolving a collision can produce another imminent
collision, so collisions are handled one-by-one and another full
collision check is run after the previous event is resolved. For a
more detailed description of the collision module in {\sc ChaNGa}, see
\citep{wallace19}.
Particles are advanced on individual timesteps chosen as a
power of two of a base timestep.  The timestep for an individual
particle is based on an estimate of the gravitational dynamical time
determined by the minimum of $\sqrt{d_{node}^3/(G(M_{node} + m_{pl}))}$
across all nodes in the tree that are accepted by the Barnes-Hut
opening criterion.  Here $d_{node}$ is the distance from the
planetesimal to the center of mass of the tree node and $M_{node}$ is
the total mass of the tree node.  For nearby particles $M_{node}$ is
replaced with the mass of the nearby particle.

\section{Narrow Annulus Simulations}\label{sec:narrow}

We begin by 
exploring the outcome of planetesimal accretion in different parts of the $\left( \alpha, \beta \right)$
parameter space. The choices of $\alpha$ and $\beta$ are motivated by two questions raised in section \ref{sec:theory}. 1) Does
runaway growth still operate when the condition that $v \ll v_{esc}$
is not satisfied? 2) How does planetesimal accretion proceed when the planetesimals themselves occupy a significant fraction of 
their Hill spheres?

To answer these questions, we run a series of simulations in which a
narrow annulus of planetesimals orbits a star. The values of $\alpha$
and $\beta$ are varied individually. 4000 planetesimals with
individual masses of $8.37 \times 10^{-5} M_{\earth}$ are placed with semimajor
axes drawn from a uniform distribution between 0.95 and 1.05 AU about a 1 $M_{\odot}$
star. In total, the disk contains $\sim$ 0.33 $M_{\earth}$ of material. The argument of perihelion, $\omega$, longitude of ascending node,
$\Omega$, and mean anomaly, M, for each body is drawn from a uniform
distribution $\in [0, 2 \pi)$. The inclinations and eccentricities are drawn
from a Rayleigh distribution with
$\left< i^{2} \right> = 1/2 \left< e^{2} \right>$ \citep{ida93a}.

In the ``fiducial'' case, we give the bodies a bulk density of 3 g
cm$^{-3}$ (for a radius of 341 km), and $\left< e^{2} \right>^{1/2} = 4 e_{h}$, which corresponds to $\alpha = 3.6 \times 10^{-2}$ and $\beta = 3.4 \times 10^{-3}$. These parameters are chosen to match the initial conditions of \citet{kokubo98}, which gave rise to oligarchic growth. 
To vary the value of $\alpha$, we alter the bulk density of the particles. In the high-$\alpha$ case, the bulk density is reduced by 
a factor of $\sim$ 7100, which produces $\alpha = 1$. This corresponds to a bulk density of $4.2 \times 10^{-4}$ g cm$^{-3}$ and a radius of 6,500 km. 
Although this is most certainly unphysical, the purpose of this modification is to have a planetesimal completely fills its Hill sphere so that no strong 
gravitational scattering should occur. To vary $\beta$, the eccentricity dispersion is increased. For the high-$
\beta$ case, $\left< e^{2} \right>^{1/2}$ is increased to $1500 e_{h}$, which corresponds to $\beta = 15,000$. This is the largest value of $\beta$ that permits a particle at 1 AU to still have an apocenter and pericenter distance that lies within the boundaries of the disk. This choice of $\beta$ places the system firmly in the $v > v_{esc}$ regime, while still allowing growth to occur in a reasonable number of timesteps.

In all cases, the simulations are evolved with a base timestep of 1.7
days, which corresponds to 3\% of an orbital dynamical time
$\sqrt{a^3/G M_{*}}$. Due to the vastly differing growth timescales in
each case, a simulation is stopped when the growth of the most massive
body flattens out. In figure \ref{fig:alpha_beta}, we show the a-e
distribution of bodies in the initial (blue) and final (orange)
snapshots from each of the 4 simulations. The size of the points
indicates the relative masses of the bodies. Only with small
$\alpha$ (case c, d) does a residual population of dynamically hot planetesimals
develop. The lack of high eccentricity planetesimals (relative to the protoplanets) in the large
$\alpha$ (case a, b) simulations suggests that most encounters result in accretion
rather than scattering. For large $\beta$ (case b, d), the growing protoplanets
and end up in a dynamically cool state,
relative with the initial conditions. This is due to kinetic energy 
being lost as particles inelastically collide. One last point we note
is the difference between the eccentricities of protoplanets in the 
large $\alpha$, large $\beta$ (case b) and the small $\alpha$,
large $\beta$ (case d) simulation. The dynamically cooler result of the former case
is likely due to the dominant role that inelastic collisions play here.

In figure \ref{fig:alpha_beta_mass}, we show the mass distribution of bodies from the final snapshot in each of the four simulations. In addition to 
leaving fewer residual planetesimals, the large $\alpha$ (case a, b) simulations produce embryos that are a factor of a few larger. Despite the 
vastly different encounter velocities of each population of bodies, the initial size of $\beta$ (so long as bodies remain in the dispersion-dominated regime) 
appears to have no significant effect on the final distribution of masses. For reference, the boundary between shear and dispersion-dominated 
encounters ($e_{h}$ = 1) lies around e = $4 \times 10^{-4}$ for the planetesimal mass we have chosen. The eccentricity at which $\left< v^{2} \right>^{1/2} = 
v_{esc}$ for the planetesimals is around $10^{-2}$.

To investigate whether any of these planetesimal rings underwent
runaway growth, we examine the time evolution of the maximum and mean
masses in each simulation. The ratio $M/\left< m \right>$ is plotted
in the top panel of figure \ref{fig:alpha_beta_evo}. Here, a positive slope
indicates that the quantities are diverging (i.e.
the growth rate is accelerating with mass). This behavior is
evident in all four cases although the small $\alpha$ simulations eventually reach a stage where the curves
 turn over as the planetesimal supply depletes. Even with a large
$\beta$, where the effective collision cross section is
 nearly equal to the geometric value, runaway growth still appears to
operate. The ubiquity of the early positive trends in this figure suggests
that as bodies collide and grow, the
relative difference in gravitational focusing factors between bodies
is what drives the system towards runaway
growth, no matter how close the collision cross sections lie to the geometric value.
Although larger encounter velocities lengthen the growth
timescales, runaway growth appears to be inevitable, so long as
gravity is the dominant force in the system. For large $\alpha$
(case a, b), the curves in this figure eventually turn over and begin to decline.
In the bottom panel of figure \ref{fig:alpha_beta_evo}, we separately show the evolution of
the maximum (solid lines) and mean (dashed lines) mass for each case. Here, it is evident
that the turnover in $M/\left< m \right>$ is
driven by an increase in the average mass as the planetesimal population
becomes depleted. For small $\alpha$, one would expect that planetesimal accretion should also eventually come to a halt as
the growth timescale lengthens due to the planetesimal surface density decreasing and the residual bodies being scattered onto high
eccentricity orbits with negligible gravitational focusing factors. Many more timesteps, however, would be required to reach this point.

Additionally, these results suggest that the value of $\alpha$, which is a function of only the initial conditions, controls the qualitative outcome of accretion. 
Across most of a planet-forming disk, $\alpha$ is small, and frequent gravitational encounters between the growing bodies will 
facilitate oligarchic growth. In the dispersion-dominated regime, close encounters drive the stirring between planetesimals and 
embryos \citep{weidenschilling89, ida90}. When $\alpha \ll 1$, the planetesimal fills only a small portion of its Hill sphere and the majority of close encounters result in viscous stirring, rather than accretion. In the opposite regime, we 
observe that runaway growth still occurs, but nearly all of the planetesimals are consumed by the forming protoplanets, rather 
than being scattered onto higher eccentricity orbits, where they would otherwise remain as a remnant of the early stages of 
planet formation \citep{kokubo98, kokubo00}.

\section{Full Disk Simulation}\label{sec:fulldisk}

\subsection{Initial Conditions}

Motivated by the qualitative dependence of accretion on $\alpha$,
we next investigate whether this highly efficient, non-oligarchic
growth should be expected to operate near the innermost regions of a typical planet-forming
disk. Given that N-body simulations of short-period terrestrial planet formation 
typically begin with a chain of neatly-spaced, isolation mass 
(see \citet{kokubo00} eq. 20) protoplanets, it is pertinent to determine 
whether the high $\alpha$ growth mode we revealed
in the previous section invalidates this choice of initial conditions.

Given the dearth of short-period terrestrial planets observed around M stars (e.g. TRAPPIST-1 \citep{gillon16, gillon17, agol21}), we chose to model the evolution of a series of wide planetesimal disks, which span from 1 to 100 days in orbital period, orbiting a late-type M star of mass 0.08 $M_{\odot}$. For a population of planetesimals with a bulk density of 3 g cm$^{-3}$, this orbital period range corresponds to $\alpha \in (0.7, 0.05)$. By simultaneously modeling a broad range of orbital periods, we can determine the critical value of $\alpha$ that divides these two modes of accretion, and also explore how the oligarchic/non-oligarchic accretion boundary affects the resulting distribution of protoplanets.

Four wide-disk simulations are run in total (see table \ref{tab:sim_properties}). In each case, the solid surface density follows a power law profile
\begin{equation}
	\Sigma(r) = \textrm{10 g cm}^{-2} \times A \left( \frac{M_{*}}{M_{\odot}} \right) \left( \frac{r}{\textrm{1 AU}} \right)^{-\delta}
\end{equation}
where $M_{*}$ is the mass of the central star, 10 g
cm$^{-2}$ is the surface density of the minimum-mass solar nebula
\citep[MMSN]{hayashi81} at 1 AU, $A$ is an enhancement factor and $\delta$ is the power law index.
In the first case (fdHi), we model a disk that follows a MMSN power law slope ($\delta$ = 1.5), with the overall normalization enhanced by a factor of 100. This choice of normalization for the solid surface density profile appears necessary in order to reproduce many observed short period terrestrial worlds in-situ \citep{hansen12}. {Many recent planetesimal formation models invoke streaming instabilities, which first require sufficiently large amounts of solid material concentrating at preferential locations in the disk. This can occur via zonal flows \citep{johansen2009b, simon12}, vortices \citep{klahr03}, or through mechanisms that produce a pressure bump in the gas disk, such as an ionization \citep{lyra08} or condensation front \citep{brauer08b, drkazowska13}, or even the perturbations from an existing planet \citep{shibaike20}. \citet{drkazowska18} showed that evolution of the snow line boundary can cause planetesimal formation over a significant area of the disk, producing mass concentrations at least as large as the ones we use here. We argue that this is a particularly attractive mechanism for widespread planetesimal formation around low-mass stars, as their extreme pre-main sequence evolution is particularly conducive to significant movement of the condensation fronts \citep{baraffe15}.

Additionally, we vary the power law index (fdHiShallow, fdHiSteep) and overall normalization (fdLo) of $\Sigma(r)$. Although there is no way to reliably measure the uncertainty on the MMSN power law slope, \citet{chiang13} applied a similar analysis to a sample of Kepler multiplanet systems and found a variation of $\sim 0.2$. Sub-millimeter observations of the outer regions of cold protoplanetary disks find that $\delta$ can be as low as 0.5 \citep{mundy00, andrews09, andrews10}. Therefore, we vary $\delta$ by a value of $1.0$ relative to the MMSN value for the fdHiShallow and fdHiSteep simulations.

In all cases, the eccentricities and inclinations of the bodies are randomly drawn from a Rayleigh distribution, with $\left< e^{2} \right>^{1/2} = 2\left<i^{2} \right>^{1/2} = e_{eq}$ \citep{ida93}. Following \citet{kokubo98}, the value of $e_{eq}$ is chosen such that the timescales for viscous stirring and aerodynamic gas drag on the planetesimals are in equilibrium. Although this approach assumes that these two mechanisms are in balance, there is nothing preventing planetesimal accretion from getting underway before the disk is sufficiently hot to be limited by gas drag. However, as we showed in the previous section, the initial dynamical state of the planetesimals does not seem to affect the outcome of accretion, so it is safe to assume that the resulting distribution of protoplanets would remain unchanged had we started with a colder disk. The viscous stirring timescale is given by \citet{ida93} as

\begin{equation}\label{eq:vs_timescale}
    t_{vs}  = \frac{\left< e^2 \right>}{d \left< e^2 \right> / dt} \approx \frac{1}{40}\left(\frac{\Omega^{2} a^{3}}{2 G m_{pl}}\right)^{2} \frac{4 m_{pl} \langle e^{2} \rangle^{2}}{\Sigma a^{2} \Omega},
\end{equation}
where $\Omega$, $a$ and $e$ are the orbital frequencies, semi-major axes and eccentricities of the individual planetesimals, respectively. In the Stokes regime, where the mean-free path of the gas is much smaller than the solid particles, the gas can be treated as a fluid and the drag timescale is given by \citet{adachi76} as

\begin{equation}\label{eq:ts_stokes}
    t_{s} = \frac{2 m_{pl}}{C_{D} \pi r_{pl}^{2} \rho_{g} v_{g}},
\end{equation}

where $C_{D}$ is a drag coefficient of order unity, $\rho_{g}$ is the local gas volume density and $v_{g}$ is the headwind velocity of the gas experienced by the planetesimal. The local gas volume density is given by
\begin{equation}\label{eq:rho_gas}
	\rho_{g} = \frac{\Sigma_{g}}{\sqrt{2 \pi} h_{g}} \exp\left[ -z^{2} / \left( 2 h_{g}^{2} \right) \right],
\end{equation}
where $\Sigma_{g}$ is the gas surface density (taken to be
240 times the solid surface density \citep{hayashi81}, $h_{g} = c_{s} / \Omega$ is the local gas scale height and $z$ is the height above the disk midplane. The sound speed profile is given by $c_{s} = \sqrt{k_{B} T(r) / \left( \mu m_{H} \right)}$, where $k_{B}$ is Boltzmann's constant, $T(r) = T_{0} \left( r / 1 \textrm{AU} \right)^{-q}$, $\mu = 2.34$ and $m_{h}$ is the mass of a hydrogen atom. For a protoplanetary disk around a typical M star, $T_{0} = 148$ K and $q$ = 0.58 \citep{andrews05}.

Finally, the headwind velocity of the gas, due to the fact that the gas disk is pressure supported, is given by

\begin{equation}\label{eq:v_gas}
	v_{g} = v_{k} \left[ 1 - \sqrt{ q c_{s}^2 / v_{k}^2} \right],
\end{equation}

\noindent where $v_{k}$ is the local Keplerian velocity (see eq. 4.30 of \citet{armitage20}). As in section
\ref{sec:narrow}, the argument of perihelion $\omega$, longitude of
ascending node $\Omega$, and mean anomaly M for the planetesimals are drawn from a uniform distribution $\in [0, 2 \pi)$.

One should note that this choice for the gas disk profile almost certainly does not capture the wide range of possibilities in real planet-forming disks. 
On one hand, a larger initial gas surface density could act to completely remove solids via radial drift, rendering in-situ accretion of solids impossible. On the 
other hand, a more tenuous gas disk might render aerodynamic drag forces completely unimportant. In this case, the random velocity of the initial 
planetesimals should be close to their mutual escape velocity. As we showed in section \ref{sec:narrow}, the initial dynamical state of the solids seems to 
have a very minimal effect on the final outcome of planetesimal accretion. In a similar vein to \citet{hansen12}, we choose to use a MMSN-like profile for the 
gas disk and instead vary the solid surface density profile to capture the range of mechanisms that might have acted to facilitate planetesimal formation in the 
first place.

\subsection{Gas Drag Force}

In addition to the mutual gravitational forces, a Stokes drag force
due to the the gas disk is applied to each particle, following the
prescription described in section 2.2.1 of \citet{morishima10}. For
the initial mass planetesimals, the Stokes number
  ($\textrm{St} = t_{s} \Omega$) at the inner and outer disk edge is
  roughly $2 \times 10^{5}$ and $10^{7}$, respectively. For $t_{s} \gg
  1$, bodies are decoupled from the gas and are only weakly affected
  by it. Because $t_{s} \sim m^{1/3}$, the Stokes number grows as planetesimal accretion proceeds, and the drag force plays an increasingly minor role. Although the aerodynamic gas drag is not expected to significantly alter the final protoplanet distribution, we include its effects here to be self-consistent with the initial conditions, which were constructed by balancing the effects of viscous stirring with gas drag.

\subsection{Timestepping Criterion}

In the case of the fdHi simulation, there are nearly 1 million
particles, whose orbital periods vary by two orders of magnitude. Because the
interaction timescales near the inner edge of the
disk are exceedingly short, a fixed timestep size would required a prohibitively large
number of steps to follow planetesimal growth throughout the entire
disk. For this reason, we use a multi-tiered timestepping scheme, in
which particles are placed onto the nearest power of two timestep
based on their most recently calculated gravitational
acceleration. This scheme is used on almost all works using ChaNGa,
and is common among large-scale simulation codes.

This more efficient scheme introduces two issues, however. Firstly,
momentum is not completely conserved when bodies switch timestep tiers. The
error introduced becomes particularly severe for a particle on an
eccentric orbit, whose perihelion and aphelion distances
straddle a timestep boundary. For a large collection of particles,
this problem manifests itself as the development of a V-shaped gap in the a-e plane, centered on the boundary itself. To correct this problem, we introduce a slightly modified timestepping criterion, which is based on the expected gravitational acceleration of the particle at pericenter. Only in the case of a close encounter with another planetesimal (in which the acceleration is no longer dominated by the star) is the timestep allowed to reduce based on the original instantaneous criterion.

A second issue is introduced when two particles on different timesteps
undergo a collision. As in the previous case, momentum is not
completely conserved because the most recent `kick' steps did not
happen simultaneously for these bodies. Early in the simulation, we
find that this problem tends to trigger runaway growth at the timestep
boundaries first. This issue carries itself forward through the embryo
formation phase, and protoplanets tend to form at the boundaries. To
correct this issue, we ignore collisions between bodies on different
timesteps early in the simulation. We find that preventing
multi-timestep collisions until after the maximum mass grows by a
factor of 10 prevents any artifacts from developing at the timestep
boundaries, while also minimizing the number of `skipped' collisions. In the case of the fdHi simulation, only about 20 collisions out of an eventual 900,000 are ignored. To verify that this timestepping scheme does not affect protoplanet growth, we tested an annulus of growing planetesimals with both fixed steps and our two-phase variable timestepping scheme. The results of these tests are shown in appendix \ref{sec:rung_ecc}.

\begin{figure}
\begin{center}
    \includegraphics[width=0.5\textwidth]{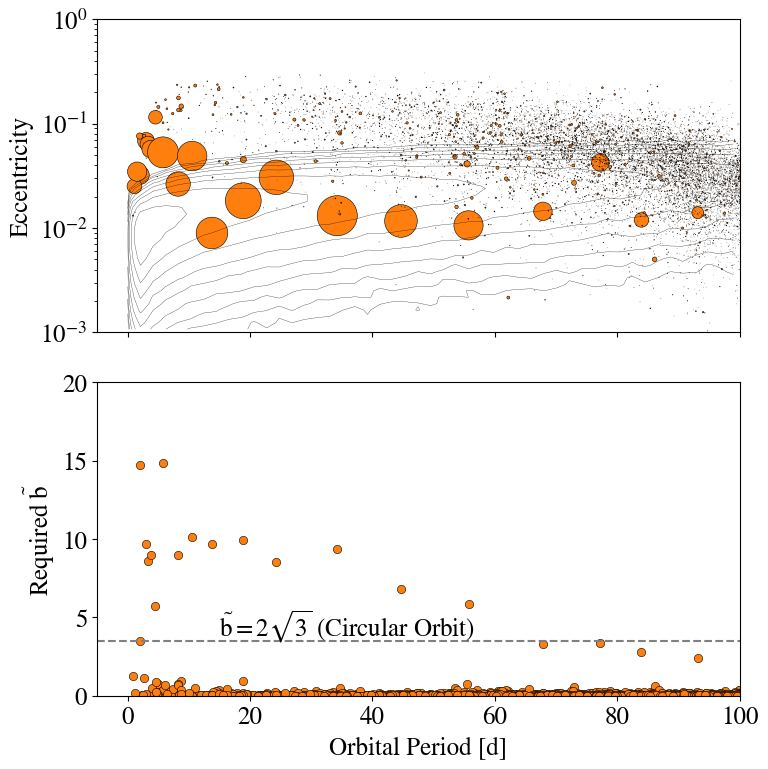}
    \caption{The final state of the fdHi simulation. In the top panel,
      the contours denote the initial period-eccentricity distribution
      of the planetesimals. Point sizes indicate the relative masses
      of bodies at the end of the simulation. In the bottom panel, we show the
      feeding zone width (see equation \ref{eq:iso}) required to produce
      the final masses of the bodies. The dashed line indicates the feeding
      zone size expected for bodies on circular orbits. For the shorter period bodies,
      the feeding zone size exceeds this expected value, which indicates that oligarchic growth is
      not operating here. This boundary occurs near roughly 60 d in orbital period.\label{fig:fulldisk_e_m}}
\end{center}
\end{figure}

\begin{figure}
\begin{center}
    \includegraphics[width=0.5\textwidth]{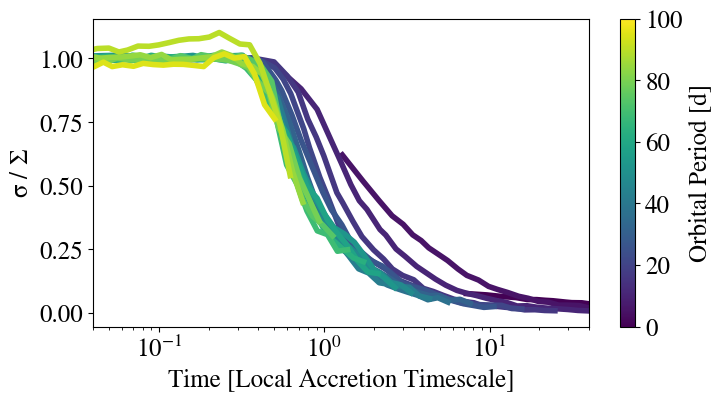}
    \caption{The time evolution of the planetesimal surface density (in units of the total solid surface density) in the fdHi 
    simulation. Each curve represents a radial slice of the disk. The time is measured in units of the local accretion 
    timescale at the center of each radial zone. The colors represent the orbital period at the center of each zone.\label{fig:pl_frac_time}}
\end{center}
\end{figure}

\begin{figure}
\begin{center}
    \includegraphics[width=0.5\textwidth]{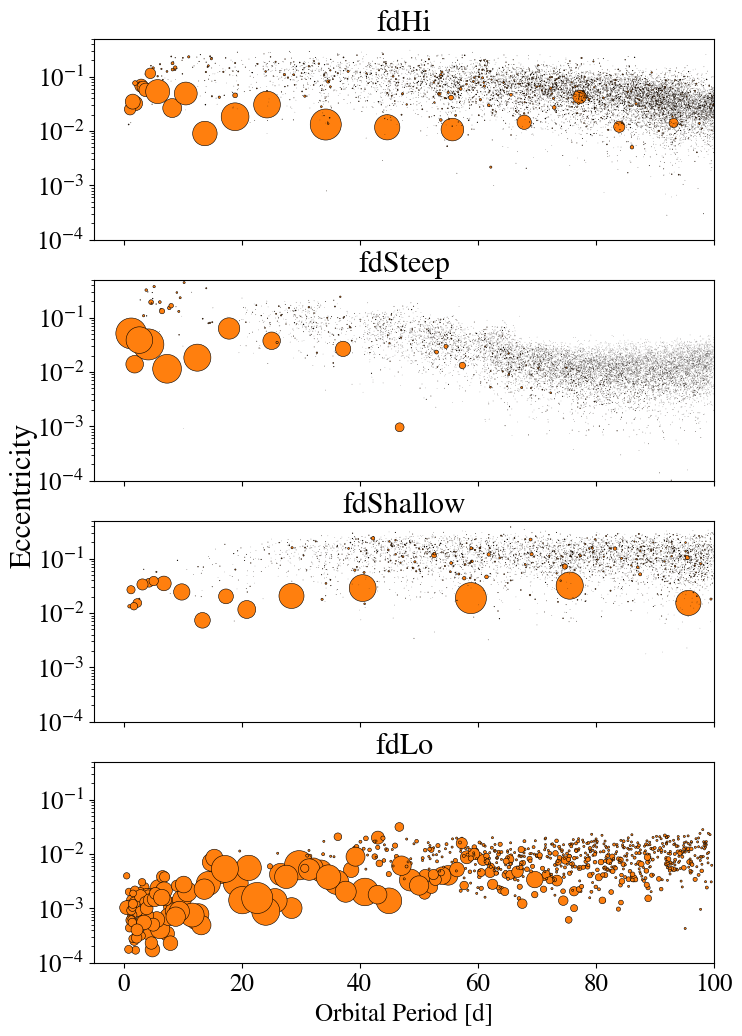}
    \caption{The final state of the full disk simulations listed in table \ref{tab:sim_properties}. 
    Point sizes indicate mass relative to the largest body in each simulation.\label{fig:surfden_profiles}}
\end{center}
\end{figure}

\begin{figure}
\begin{center}
    \includegraphics[width=0.5\textwidth]{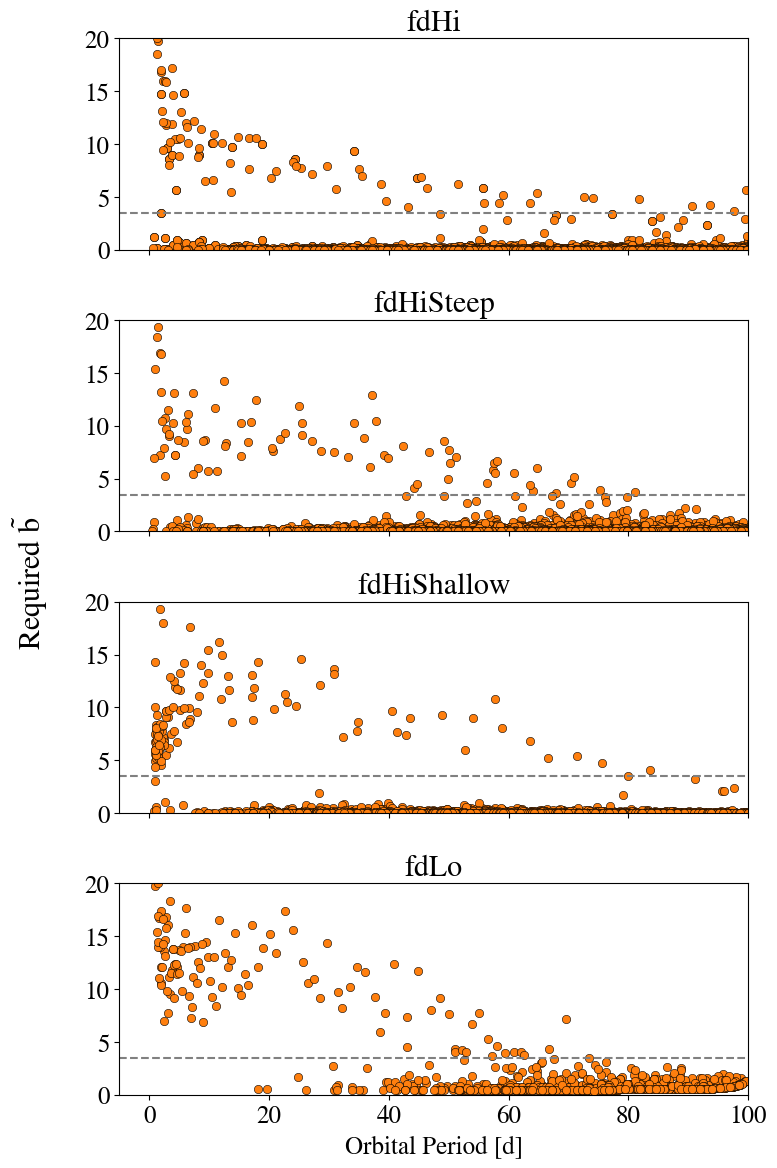}
    \caption{Feeding zone width (see equation \ref{eq:iso}) required to produce the final masses for the protoplanets from the 
    simulations listed in table \ref{tab:sim_properties}. For the fdHi, fdHiSteep and fdHiShallow simulations, we have included the results from five separate iterations of the simulations, each using a different random number seed. The horizontal dashed line indicates $\tilde{b} = 2 \sqrt{3}$. Despite the 
    vastly different initial solid surface density profiles, the feeding zone width reaches the circular orbit value around $\sim$ 60 
    days in all cases. \label{fig:surfden_b}}
\end{center}
\end{figure}

\subsection{Results}

\begin{table*}
\begin{center}
\caption{Summary of Full Disk Simulations}
\begin{tabularx}{1.0\textwidth}{l@{\extracolsep{\fill}}ccccccc} \hline \hline
Simulation Name & $m_{pl} [M_{\earth}]$ &$N_{pl}$ & A & $\delta$ & $M_{PP}  [M_{\earth}]$ & $T_{int} [yr]$ & $T_{int, 1} [yr]$ \\ \hline
fdHi              & $8.37 \times 10^{-6}$ & 903,687 & 100 & 1.5 & 1.00 & 456 & 16,377  \\
fdHiSteep    & $8.37 \times 10^{-6}$ & 903,687 & 100 & 2.5 & 1.19 & 456 & 16,377  \\
fdHiShallow & $8.37 \times 10^{-6}$ & 903,687 & 100 & 0.5 & 1.08 & 456 & 16,377 \\
fdLo             & $8.37 \times 10^{-6}$ & 45,185  & 1      & 1.5 & $1.77 \times 10^{-3}$ & 3,713 & 133,651 \\ \hline
\end{tabularx}\\
\begin{flushleft}
\textit{Note:} A summary of the four `full disk' simulations presented in section \ref{sec:fulldisk}. $m_{pl}$ and $N_{pl}$ are the initial masses and number of planetesimals. $A$ and $d$ is the initial power law normalization and slope of the solids in the disk. $M_{PP}$ is the maximum protoplanet mass at the end of the simulation and $T_{int}$ is the amount of time each simulation was integrated. $T_{int, 1}$ is the integration time scaled by a factor of $f^{2}$, which accounts for the fact that the accretion timescale has been shortened by the inflated collision cross sections.
\end{flushleft}
\label{tab:sim_properties}
\end{center}
\end{table*}

The timescales for embryo formation depend on
the chosen surface density profile, along with the local orbital
timescale. Protoplanets form first at the inner edge of the disk,
where the dynamical timescales are short. Growth proceeds in an
inside-out fashion, with the outermost regions of the disk completing
the protoplanet assembly phase last (as an example, see figure 1 of \citet{kokubo02}. This radial timescale dependence is not typically
accounted for in planet formation simulations a notable exception being \citet{emsenhuber21a, emsenhuber21b}, and appears to be an
important component to forming realistic solar system analogs
\citep{clement20}. As with the narrow annulus simulations, we stop the
integration once the masses of protoplanets in the outermost region of
the disk reach a steady value. In table \ref{tab:sim_properties}, we
summarize the outcomes of the four ``full disk'' cases.

We show the final state of the ``fdHi'' simulation in figure \ref{fig:fulldisk_e_m}. In the top panel,
the initial (contours) and final (points) state of the simulation is
shown in the orbital period-eccentricity plane. The size of the points
indicates the relative mass of the bodies. In the bottom panel, the
final masses of the bodies (in units of feeding zone size $\tilde{b}$) are shown
as a function of orbital period. The y-values in the bottom panel of figure \ref{fig:fulldisk_e_m} are calculated by solving equation \ref{eq:iso} for $\tilde{b}$, and inputting the initial surface density and final particle mass into the expression. In other words, $\tilde{b}$ is describing the size of the annulus that must be cut out of the planetesimal disk in order to produce a protoplanet of the current mass. By plotting the derived value of $
\tilde{b}$ as a function of orbital period, differences in the dynamical interactions at different locations of the disk are made more 
clearly visible. The feeding zone size $\tilde{b} = 2 \sqrt{3}$ permitted by bodies on circular, non-inclined orbits \citep{naka88} is shown by the
horizontal dashed line. In typical oligarchic growth simulations \citep{kokubo98}, protoplanets tend 
to space themselves apart by $\tilde{b} = 10$, although it should be noted that they do not consume all of the planetesimals 
within this distance.

A qualitative shift in the protoplanet and planetesimal distribution is visible inside of $\sim$ 60 days. Interior to this location, there are 
very few remaining planetesimals and the embryos formed have larger feeding zones. Exterior to the boundary, the residual 
planetesimal population is much more visible, and protoplanets more closely follow the $\tilde{b} = 2 \sqrt{3}$ line. This 
suggests that the transition between the low $\alpha$ and high $\alpha$ accretion modes seen in section \ref{sec:narrow} 
is happening near this location.

In section \ref{sec:narrow}, we postulated that the increased importance of inelastic damping in the inner, non-oligarchic growth 
region of the disk should lower the overall eccentricity of the protoplanets there. This behavior is not immediately apparent in the 
top panel of figure \ref{fig:fulldisk_e_m}. In fact, the opposite appears to be true. There are, however, a couple of factors in the wide disk simulations that could make this 
extra dynamical cooling mechanism difficult to see. Firstly, the initial eccentricity distributions of the inner and outer disk are 
different because of the dependence of the viscous stirring and gas drag timescales on orbital period. The mean eccentricity at the outer disk edge is 4x larger than at the inner disk edge. Additionally, the 
protoplanet formation timescales for the inner and outer disk are vastly different, making a comparison between these regions at 
the same moment in time somewhat inappropriate. A quick back of the envelope calculation yields $\left< e^2 \right>^{1/2} = 0.05$ for a population of $\sim 1 M_{\earth}$ bodies with a random velocity dispersion equal to their mutual escape velocity. It is therefore likely the case that the innermost protoplanets have had ample time to self-stir.

To ensure that the boundary seen around 60 days in orbital period is not
simply a transient product of the inside-out growth throughout the
disk, we examine the time evolution of $\sigma/\Sigma$, which compares the planetesimal and total solid surface density at 
multiple orbital periods. In figure \ref{fig:pl_frac_time}, the value of
$\sigma/\Sigma$ is plotted as a function of time in 10 orbital period
bins, each with a width of 10 days. To determine whether the evolution of the planetesimal surface density behaves self-similarly across the disk,
we normalize the time values in each bin by the local accretion timescale at the beginning of the simulation, which is given by

\begin{equation}\label{eq:acctime}
	t_{acc} = \left( n \Gamma v \right)^{-1} = \left( \frac{\Sigma_{0} \Omega}{2 m_{pl}} \Gamma \right)^{-1},
\end{equation}

\noindent where we have assumed the local number density of particles n is set by the surface density and the local scale height of the disk (see section \ref{sec:theory}). The effective collision cross section is set by gravitational focusing and is given by equation \ref{eq:gravfocus}.

The color of the curves indicate the orbital period bin which is being measured. From about 40 to 100 days in orbital period, the planetesimal surface 
density follows a similar trajectory as accretion proceeds. Interior to about 40 days, $\sigma$ actually decays more slowly. In other words, growth is actually 
fueled less vigorously by planetesimals in this region. This highlights the fact that accretion proceeds in a qualitatively different way in the inner disk. For the 
outer disk, gravitational focusing tends to facilitate collisions between protoplanets and preferentially smaller bodies. At short period, however, all close 
encounters result in a collision, regardless of mass. In a rather counterintuitive fashion, planetesimals in the inner disk actually persist for longer. In section 
\ref{sec:assembly}, we examine the assembly history of the embryos and show that there is much less of a preference for planetesimal-embryo collisions at 
short period as well.

In the inner disk, this value asymptotes to zero as the 
planetesimal population entirely depletes. In the outer disk, dynamical friction between the embryos and planetesimals 
eventually throttles subsequent accretion and leaves $\sim$ 10 percent or more of the mass surface density as planetesimals. It 
should be noted that in a typical oligarchic growth scenario, where protoplanets space themselves apart by 10 $r_{h}$ and settle 
onto circular orbits (giving $\tilde{b} = 2 \sqrt{3}$), roughly 30 percent ($2\sqrt{3}/10 \simeq 0.3$) of the planetesimals
should remain out of reach of the protoplanets.

Next, we investigate how the resulting planetesimal and protoplanet distribution
changes as we vary the initial solid surface density profile.
The final orbital period-eccentricity state of the
particles in the fdHi, fdHiSteep, fdHiShallow and fdLo simulations are shown in figure \ref{fig:surfden_profiles}, with point sizes 
indicating the relative masses of the bodies. In all cases, the inner disk is largely depleted of planetesimals, while the outer disk 
contains a bimodal population of planetesimals and embryos, with a clear separation in eccentricity between the two. Despite 
having significantly different masses, the semimajor axis-eccentricity distributions of the planetary embryos formed in all simulations are remarkably similar. This 
is likely due to the fact that inelastic collisions play a more significant role where the solid surface density is highest, which 
offsets the fact that the initial bodies started off in a dynamically hotter state (due to the increased effectiveness of viscous stirring). The only exception to this is the fdLo simulation, 
where the resulting eccentricities are a couple orders of magnitude smaller. Inelastic damping likely plays an even more 
significant role here, due to the much larger masses of the initial planetesimals.

In figure \ref{fig:surfden_b}, we plot the masses of the resulting protoplanets and planetesimals in all four simulations in units of $
\tilde{b}$ (see equation \ref{eq:iso}). To make the trend between $\tilde{b}$ and orbital period more clear, we ran four more versions of the fdHi, fdHiSteep and fdHiShallow simulations using different random number seeds and included these in the figure as well.
As mentioned previously, $\tilde{b} = 2 \sqrt{3}$ (indicated by the horizontal dashed line) is the  
feeding zone width that a body on a circular orbit will have. In all four simulations, the feeding zone width exceeds the 
minimum value in the inner disk and approaches $2 \sqrt{3}$ around $\sim$ 40 to 60 days. The orbital periods at which this transition 
occurs are quite similar between simulations, despite the vastly different solid surface density profiles used. This indicates that 
the boundary between accretion modes is driven entirely by the local value of $\alpha$, and also supports our 
conclusion that planetesimal accretion is largely complete everywhere in the disk.

\subsection{Assembly History of Embryos}\label{sec:assembly}

Further insight regarding the difference between the short vs long period accretion modes can be gained by looking at the 
growth history of the planetary embryos. Because all collisions are directly resolved by the N-body code, a lineage can be 
traced between each planetary embryo and the initial
planetesimals. For the fdHi, fdHiSteep and fdHiShallow
  simulations, protoplanets gain a factor of $\sim 10^6$ in mass
  relative to the initial planetesimals. For the fdLo simulation, this growth factor is nearly a thousand times smaller, which produces rather shallow and noisy collision histories. For this reason, we choose to exclude the fdLo simulation from our analysis in this section.

We begin by investigating the ``smoothness'' of the accretion events
that give rise to each embryo. Drawing from a common technique used
for cosmological simulations of galaxy formation, we divide growth events for
a given body into ``major'' and ``minor'' mergers \citep{kauffmann93, murali02, lhullier12}. 
Here, we define minor events as any collision involving an initial mass 
planetesimal, while major events consist of a merger between any two larger bodies.
In figure \ref{fig:minor_frac}, we retrieve the collision events for all bodies in all five iterations of the fdHi, fdHiSteep and fdHiShallow
simulations and plot the total mass fraction attained through minor merger (smooth accretion) events as a function of the final mass of the body. Here, we define a minor merger
to be any collision involving a planetesimal with $m < 100 m_{0}$. The color of the
points indicates the final orbital period of the body. Beyond $\sim$ 20 to 30 days in orbital period, minor mergers make up a significant fraction of the final mass of a body. Interior to this, the smooth accretion fraction drops significantly and the mass contribution to minor mergers can vary by over an order of magnitude.

\begin{figure}
\begin{center}
    \includegraphics[width=0.5\textwidth]{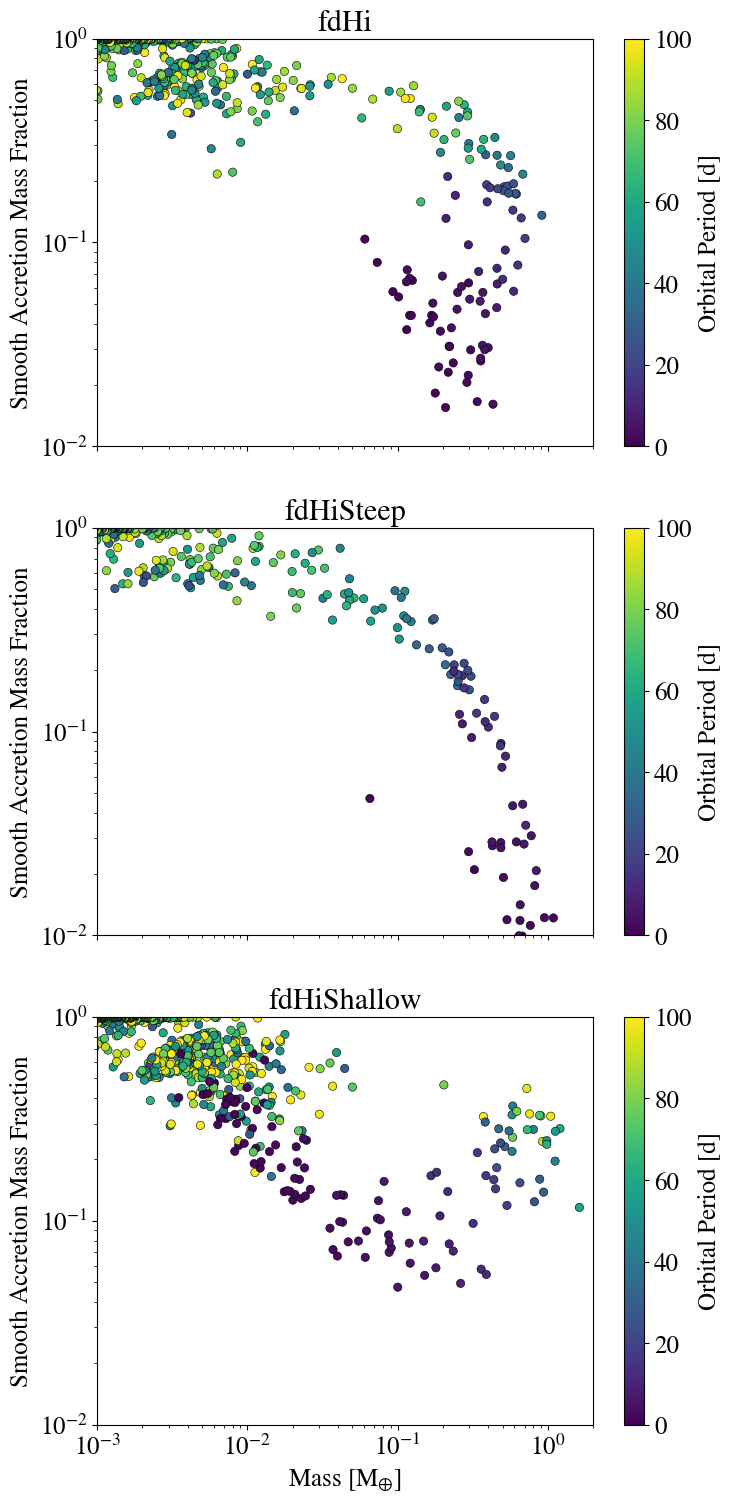}
    \caption{For all bodies with $m > 100 m_{0}$ at the end of the high surface density simulations, the fraction of their total mass attained through mergers with initial mass planetesimals (smooth accretion) as a function of total mass. Colors indicate the orbital periods of the bodies in the final simulation snapshot. Bodies interior to $\sim$ 20 to 30 days attain up to an order of magnitude less of their mass through minor merger events, while accretion of planetesimals plays a much more significant role at longer orbital periods.\label{fig:minor_frac}}
\end{center}
\end{figure}

\begin{figure}
\begin{center}
    \includegraphics[width=0.5\textwidth]{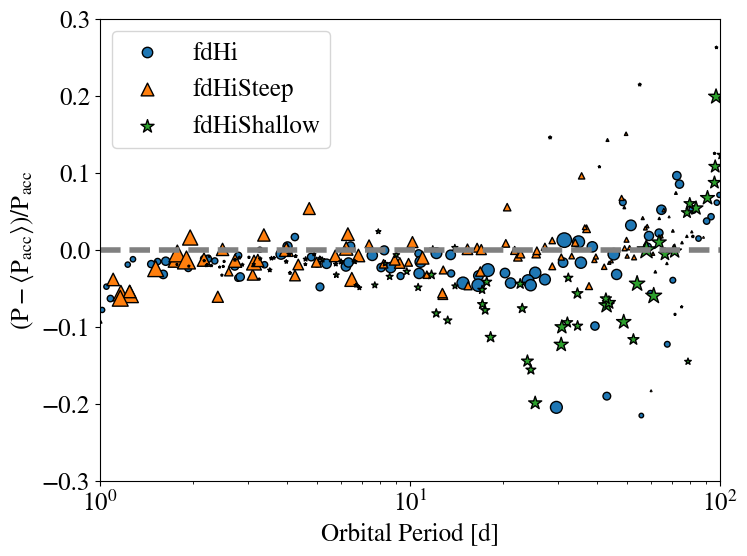}
    \caption{The relative separation between the final period
        of a body and the mean orbital periods of its accretion zone at the end of the fdHi, fdHiSteep and fdHiShallow simulations. The marker type and color denotes the simulation used, while the marker sizes indicate the relative masses of bodies. In each case, all five iterations of the simulations are plotted simultaneously.\label{fig:acc_zones}}
\end{center}
\end{figure}

The variation in smooth accretion fraction with mass for the short period bodies suggests that the planetesimal and embryo populations interact differently than those in the outer disk. Exterior to the accretion mode boundary, the growing embryos continue to accrete planetesimals while avoiding each other as they near their final mass. Inside the boundary, however, any and all bodies collide with each other, and the occasional embryo-embryo collision tends to dominate growth and drive down the smooth accretion fraction. Gravitational scattering between embryos and planetesimals is a key ingredient for orbital repulsion \citep{kokubo98}, and so a lack of gravitational scattering in the inner disk should prevent the embryos from settling onto neatly-spaced, isolated orbits. As we showed in figure \ref{fig:surfden_b}, the embryos in the inner disk appear to reach well beyond the typical feeding zone size predicted by an oligarchic growth model. Figure \ref{fig:minor_frac} suggests that the extra mass here obtain comes from mergers with the other nearby embryos.

Another line of evidence pointing to a lack of gravitational
scattering and orbital repulsion in the inner disk can be seen in
figure \ref{fig:acc_zones}. Here, we measure the initial orbital period distribution of bodies used to construct each embryo and calculate its mean $\left<  P_{acc} \right>$. Point sizes indicate the relative final masses of bodies. As in the previous figure, we have included data from all five versions of the fdHi, fdHiSteep and fdHiShallow simulations. For each body, this quantity is then compared with its final orbital period. In this figure, bodies that closely follow the gray dashed line still reside close to their initial feeding zones. On the other hand, bodies further from the dashed line must have experienced a strong gravitational scattering event during or after their accretion has completed. For all of the high surface density simulations, the accretion zones and present positions of the embryos appear to diverge beyond $\sim$ 20 days in orbital period.

Coupled with the strong decrease in smooth accretion fraction for bodies in this region (seen in figure \ref{fig:minor_frac}), it appears as if the relative importance of collisions and gravitational scattering seems to shift around $\sim$ 20 days. For the shortest period bodies, growth events are sudden and stochastic, often involving collisions between bodies of comparable mass. For longer period bodies, a significant amount of growth is driven by accretion of smaller planetesimals. We postulate that this qualitative difference is driven by the role that embryo-embryo close encounters play in the inner and outer disk. In the inner disk, these encounters tend to result in a merger, which drives down the smooth accretion fraction. In the outer disk, these encounters tend to result in a scattering event which moves bodies away from their initial feeding zones. We find that the accretion zone shapes for the longer period bodies are also much more smooth and unimodal, which suggests that scattering tends to occur after accretion has largely completed.

\section{Simplifying Assumptions}\label{sec:assump}

\subsection{Collision Cross Section}

In all cases shown so far, the boundary between the
oligarchic growth and the highly-efficient short period accretion
region lies between 40 and 70 days in orbital period. As discussed in section \ref{sec:sizeandhill}, the mode of accretion is
set entirely by the local value of $\alpha$, which scales with both
distance from the star and the bulk density of the planetesimals (see
equation \ref{eq:alpha}). Because we chose to artificially inflate the
collision cross section of the particles in our simulations by a factor of $f$, the bulk densities
of the particles are reduced, and the accretion boundary is shifted outward.
However, the scaling relation between $\alpha$ and $\rho$ ($\alpha \sim \rho_{pl}^{-1/3}$) can be used
to predict where this accretion boundary should lie in a disk with realistic-sized planetesimals. The simulations
presented in this paper use a collision cross section enhancement factor of 6, which moves the boundary outward in orbital
period by a factor of approximately 15 (For a fixed value of $\alpha$, equation \ref{eq:alpha} gives $a \sim r_{pl} \sim f$ and therefore $P_{orbit} \sim f^{-3/2}$). One would therefore expect the accretion boundary to lie between 3 and 5 days in orbital period
for 3 g $cm^{-3}$ bodies.

Although a simulation with $f=1$ is not computationally tractable, we
can test whether the accretion boundary moves in the way we expect by
modestly changing the value of $f$. In figure \ref{fig:f6f4_b}, we
compare the fdHi simulation to a nearly identical run using $f=4$. In
the top panel, we show the feeding zone width required for each particle to
attain its present mass. As in figure \ref{fig:surfden_b}, we indicate the feeding
zone size expected for oligarchic growth with a horizontal dashed line. In the bottom panel, the value
of $\alpha$ as a function of orbital period is shown for 3 g $cm^{-3}$
bodies with an artificial radius enhancement of $f=1$, 4 and 6. The
horizontal dashed line indicates the empirical value of alpha below which the
accretion mode switches to oligarchic. Comparing the top and
bottom panels, the intersection of the feeding zone width seen in our simulations and the
feeding zone width predicted by oligarchic growth matches well with the orbital period at which $\alpha
\sim 0.1$ for both values of $f$.  Also shown by the shaded region are the expected $\alpha$ values for
realistic-sized bodies with $\rho_{pl}$ between 1 and 10 g cm$^{-3}$. Although the removal of the cross section
enhancement greatly reduces the size of the non-oligarchic region, it still should be expected to cover a portion of the
disk where planetesimals might be expected to form \citep{mulders18} for a wide range of $\rho_{pl}$.

\begin{figure}
\begin{center}
    \includegraphics[width=0.5\textwidth]{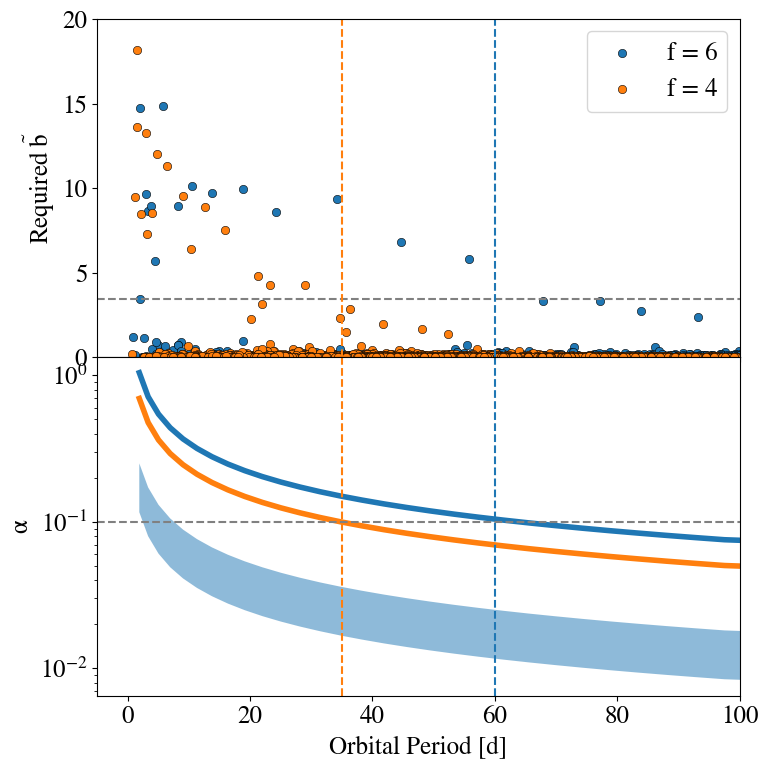}
    \caption{In the top panel, we show the required feeding zone sizes to produce the masses of the bodies seen
    at the end of the fdHi and fdHif4 simulations.  The bottom panel shows the variation of $\alpha$ with orbital period for the 
    bodies used in each case (solid curves). The orbital period at which $\alpha \simeq 0.1$ matches well with the location at 
    which $\tilde{b}$ exceeds $2 \sqrt{3}$ This is highlighted by the vertical dashed lines. The shaded region
    in the bottom panel show the values of alpha for realistic-sized ($f=1$) planetesimals with bulk densities between 10 and 1 g cm$^{-3}$.\label{fig:f6f4_b}}
\end{center}
\end{figure}

\subsection{Collision Model}

\begin{figure}
\begin{center}
    \includegraphics[width=0.5\textwidth]{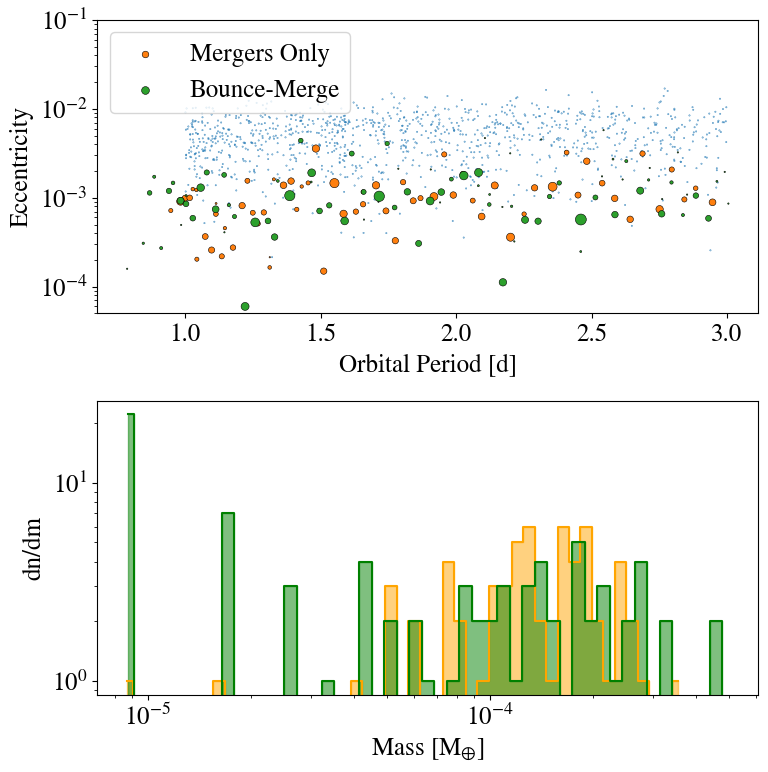}
    \caption{A comparison between the innermost region of the fdLo (orange) simulation, and a second version using a bounce-
    merge collision model (green). In the top panel, the period-eccentricity state of the particles is shown, with marker sizes 
    indicating relative mass. The blue points represent the initial state of the simulations. The bottom panel compares the final 
    differential mass distributions of the bodies. \label{fig:frag_ecc}}
\end{center}
\end{figure}

\begin{figure}
\begin{center}
    \includegraphics[width=0.5\textwidth]{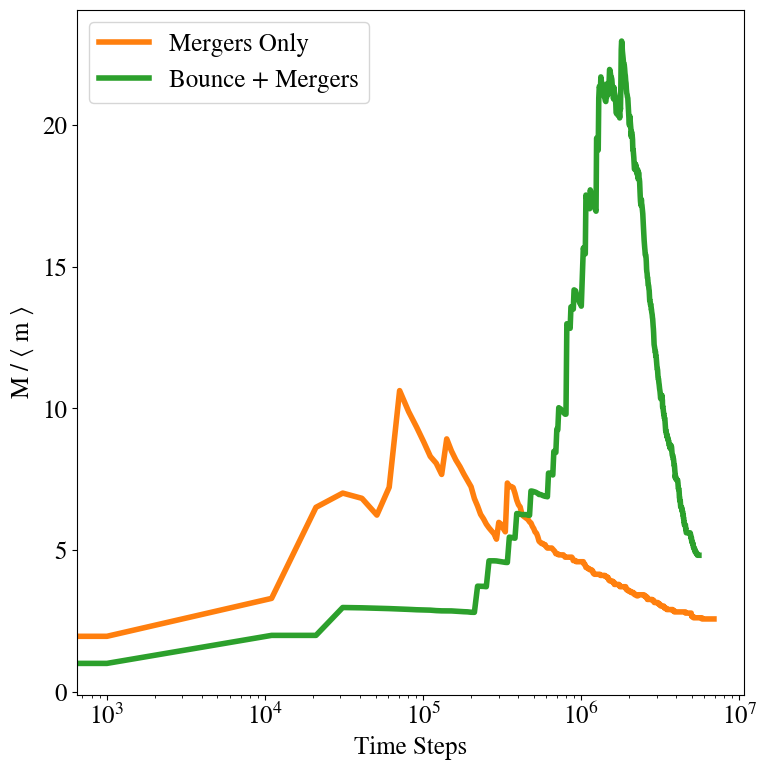}
    \caption{The evolution of the ratio between the maximum and mean mass of the simulations shown in figure \ref{fig:frag_ecc}. 
    In both cases, the system first evolves through a phase of runaway growth, before the massive bodies consume the smaller 
    bodies, driving down the mean mass. With the bounce-merge model, the mass ratio takes much longer to begin decreasing.\label{fig:frag_evo}}
\end{center}
\end{figure}

For the simulations presented in this work, every collision results in a perfect merger between pairs of bodies, with no loss
of mass or energy. Although simpler and less computationally expensive to model, allowing every collision to produce a
perfect merger might result in overly efficient growth, particularly in the innermost region of the disk where the encounter 
velocities are largest. Given that we have just shown that a distinctly non-oligarchic growth mode emerges in the inner disk when the 
collision timescale is short relative to the gravitational scattering timescale, one might be concerned that a more realistic collision model would act 
to lengthen the growth timescale enough for this condition to no longer be true. In the outer regions of the disk where oligarchic growth still 
operates, more realistic collision models have been shown to simply lengthen the timescale for planetary embryos to form 
\citep{wetherill93, leinhardt05}. A proper way to handle this would be to allow for a range of collision outcomes, based on a semianalytic 
model (see \citet{leinhardt12}). However, resolving collisional debris, or even prolonging growth by forcing high-velocity pairs of 
bodies to bounce is too expensive to model, even with {\sc ChaNGa}.

To test whether a more restrictive collision model should alter the growth mode of the inner disk, we ran a smaller scale test 
using a more restrictive collision model. In this case, a collision can result in one of two outcomes: if the impact velocity is 
smaller than the mutual escape velocity of the colliding particles, defined as
\begin{equation}\label{eq:v_mut}
	v_{mut, esc} = \sqrt{\frac{2 G (m_{1} + m_{2})}{r_{1} + r_{2}}},
\end{equation}
where $m_{1}, m_{2}$ and $r_{1}, r_{2}$ are the masses and radii of
colliding particles 1 and 2, then the bodies merge. For
impact velocities larger than $v_{mut, esc}$, no mass is transferred, and the bodies undergo a completely elastic bounce. 
Because the accretion outcome is all or nothing, this model should restrict growth more than a partial accretion model 
\citep{leinhardt12}. Below, we will show that the bounce-merge model does not meaningfully affect the outcome of the inner 
disk's planetesimal accretion phase, and so a more realistic partial accretion model should do the same.

To compare the outcome of the two collision models, we have chosen to use the initial conditions from the fdLo simulation, but 
have truncated the disk beyond 3 days in orbital period. This offsets the increased computational cost of the more restrictive collision model, while still allowing the disk to evolve in the region where mergers would be most difficult to achieve.
For the initial conditions we have chosen, the typical encounter velocity (defined by $v_{enc} = \left< e^{2} \right>^{1/2} v_{k}$, 
where $v_{k}$ is the local Keplerian velocity) is about 25 percent larger than $v_{mut, esc}$. Because the encounter velocities 
follow a Gaussian distribution, there should still be a small subset of collisions that still meet the merger criteria to occur early on. 
In addition, $v_{mut, esc}$ becomes larger as the bodies grow and the merger criteria should become easier to meet as the 
system evolves. For these reasons, one would expect the inhibition of growth due to the more restrictive collision model to be 
temporary.

In figure \ref{fig:frag_ecc}, we compare the outcomes of the simulations, one with mergers only (shown in orange) and one with 
the bounce-merge model (shown in green). The blue points in the top panel show the initial conditions used for both cases. 
Although the bounce-merge simulation takes much longer to reach the same phase of evolution, the resulting orbital properties
are indistinguishable from the merger-only case. Performing a Kolmogrov-Smirnov test on the two mass distributions yields
a p-value of $2 \times 10^{-5}$, which tells us that the two mass distributions are quite firmly statistically different. If we remove
the initial mass planetesimals, a KS test yields a p-value of 0.1, which suggests that the distributions are statistically similar. Because the remaining planetesimals only make up about 0.1 percent of the total mass of the disk, we conclude that the embryo populations are nearly indistinguishable, while the bounce-merge model produces a small amount of residual planetesimals.

To investigate the differences in growth between the two collision models early on, we show the time evolution of the ratio 
between the maximum and mean mass in figure \ref{fig:frag_evo}. In both cases, this ratio first increases, which indicates that 
runaway growth still operates, regardless of the collision model used. In the bounce-merge case, the mass ratio peaks at a 
higher value, while also undergoing a longer runaway growth phase. This suggests that the mass distribution becomes much 
less unimodal during this growth process, but as figure \ref{fig:frag_ecc} shows, this does not affect the resulting embryos or 
allow for a residual planetesimal population.

As a final note, \citet{childs22} found that a more realistic collision model also enhanced radial mixing in their simulations. Upon 
calculating the planetesimal accretion zones using the same method as was done to produce figure \ref{fig:acc_zones}, we find 
that the embryos in the bounce-merge simulation annulus do have modestly wider accretion zones than those produced in the 
merger-only simulation.

\section{Summary and Discussion} \label{sec:discuss}

In this work, we have demonstrated that planetary embryo growth
can simultaneously operate in two distinct modes in a planet-forming disk. In the first
mode, gravitational feedback from the growing embryos heats the
remaining planetesimals and results in a dynamically cold population
of embryos with a modest amount of residual planetesimals. This
corresponds to the ``oligarchic growth'' case revealed by \citep{kokubo98}, which is often used as a starting point for late-stage 
accretion models (e.g. \citet{kokubo02, raymond05, raymond06}). In the second mode, the gravitational feedback does not 
play a significant role, embryos quickly sweep up all planetesimals, and grow larger and less uniformly spaced than those produced by 
oligarchic growth.

We have demonstrated the outcome of both accretion modes through a
simple parameter study using a narrow annulus of planetesimals (section \ref{sec:narrow}). The initial planetesimal distribution can be described in terms of two dimensionless 
constants, $\alpha$ and $\beta$, which describe the ratio between the physical radius of the planetesimals and the Hill ($r_{h}$) 
and gravitational ($r_{g}$) radius, respectively. For a fixed planetesimal composition, $\alpha$ scales with the orbital period and 
$\beta$ scales with the level of dynamical excitation of the disk. We showed that $\alpha \ll 1$ leads to oligarchic growth, while an $\alpha$ close to unity produces this newly revealed non-oligarchic growth mode (see figure \ref{fig:alpha_beta}). Within this non-oligarchic mode, we find that 
the resulting masses and eccentricities of the embryos come out very similar, regardless of the initial value of $\beta$.

So long as the density of the bodies do not significantly change as
their mass distribution evolves, this ratio is set entirely by the distance
from the star. Because both the physical and Hill radii of the bodies
grows as $M^{1/3}$, the growth mode boundary remains stationary
 in the disk during the planetesimal accretion process.
 
We have verified that the growth boundary location does not strongly depend of the
solid surface density distribution by testing the outcome of the planetesimal
accretion process for a variety of solid
profiles. Although altering the surface density does affect the
resulting masses of the embryos, the location of the boundary
separating the growth modes is remarkably similar among all of our simulations.
In addition, the sizes of the feeding zones, along with qualitative differences in the
accretion history of embryos on both sides of the boundary (see figures \ref{fig:minor_frac} and \ref{fig:acc_zones}) provide further evidence to
suggest that oligarchic growth is not operating in the inner disk.

Finally, we have examined how our assumption of perfect accretion, along with the collision cross section enhancement used, might alter our results. 
We verified that these modifications, meant to make the simulations less computationally expensive, would still allow for the emergence of this 
non-oligarchic growth mode.
We showed that a much more restrictive collision model, in which only low-velocity collisions produce a merger, still allows for 
this growth mode to occur at the innermost part of the disk, where encounter speeds are most vigorous. In a real planet-forming 
disk, partial accretion events should allow growth to happen more quickly than what was seen in this test case (see figure 8 of \citet{leinhardt15}), so this growth 
mode should certainly still occur. We also showed that the collision cross section enhancement moves the accretion boundary 
outward. We verified this by deriving a scaling relation between the boundary location and the bulk density of the planetesimals, 
and showing that the boundary moves to the predicted location when running a simulation with a slightly smaller inflation factor. 
For rocky planetesimals with a realistic bulk density, 3 g cm$^{-3}$, our results suggest that this boundary should lie around 5 days in orbital period.

\subsection{Connections to Satellitesimal Accretion}

To date, there have been no other studies of planetesimal accretion
with such a large value of $\alpha$. Typically, it is assumed that $\alpha \ll 1$ (e.g. \citet{lithwick14}), which is certainly true for material at and beyond the Earth's orbit. However, a value of $\alpha = 1$
corresponds to the Roche limit of a three-body system, and so one
might wonder this high-$\alpha$ accretion mode might be relevant for a
circumplanetary accretion. There is a collection of previous works
which use N-body methods to examine in-situ satellitesimal accretion
\citep{ida97, richardson00, kokubo00b, ida20}, although some of these simulations involve a complex interaction between spiral 
density waves formed inside of the Roche limit and the material exterior to it, making the dynamics driving accretion distinctly non-local, in contrast to what we have presented in this work. \citet{ida97} was able to form 1-2
large moons just exterior to the Roche limit, depending on the extent of the disk with very little satellitesimal material left over. 
The widest disk they modeled extended out to $\alpha = 0.5$. Qualitatively, this result is very similar to the short period 
planetesimal accretion mode observed in our simulations. \citet{ida20} modeled a much wider satellitesimal disk, which extends 
out to about $\alpha \approx 0.05$. Inside to the $\alpha = 0.1$ accretion boundary (which lies near 15 planetary radii in figure 1 of 
\citet{ida20}), bodies grow beyond the isolation mass, while the opposite is true on the other side of the boundary. In addition, a 
residual population of satellitesimals is still present beyond the boundary, which suggests that oligarchic growth is indeed 
operating only on the far side.

Presently, the implications that this non-oligarchic accretion mode has for the formation of short-period terrestrial planets, and 
whether the accretion boundary would leave any lasting imprint on the final orbital architecture, is unclear. The extreme efficiency of planetesimal accretion at the inner edge of the disk suggests that no residual populations of small bodies should be expected to exist here. A crucial point that our 
results do highlight is that the initial conditions used for most late-stage planet formation simulations are overly simplistic. 
\citet{clement20} recently simulated planetesimal accretion in a disk extending from the orbit of Mercury to the asteroid belt and 
found that the disk never reaches a state in which equally-spaced, isolation mass embryos are present everywhere 
simultaneously. Instead, different annuli reach a `giant impact' phase at different times, preventing the onset of a global instability 
throughout the entire disk, as is common in classic terrestrial planet formation models \citep{chambers01, raymond09}.

To connect these accretion modes to the final orbital architecture,
and to ultimately determine what implications an in-situ formation model has for the growth of STIPs, we will evolve the final 
simulation snapshots presented here with a hybrid-symplectic integrator for Myr timescales. The final distribution of planets 
formed, along with composition predictions generated by applying cosmochemical models to our initial planetesimal distributions 
and propagating compositions through the collision trees, will be examined in a follow-up paper.

\section*{Acknowledgements}
We would like to thank the anonomyous referee for their thorough and careful comments which greatly improved the quality of this manuscript. This work used the Extreme Science and Engineering Discovery Environment (XSEDE), which is supported by National Science Foundation grant number ACI-1548562. SCW and TRQ were were supported by National Science Foundation grant number AST-2006752. We acknowledge the people of the Dkhw’Duw’Absh, the Duwamish Tribe, the Muckleshoot Tribe, and other tribes on whose traditional lands we have performed this work.

\textit{Software:} Astropy \citep{astropy13}, {\sc ChaNGa} \citep{jetley08, menon15}, Matplotlib 
\citep{matplotlib07}, NumPy \citep{numpy11}, Pandas \citep{pandas10}, {\sc PYNBODY} \citep{pynbody13}

\section*{Data Availability}
The data presented in this article, along with a python script to generate the figures are available at at \href{https://doi.org/10.5281/zenodo.8140487}{10.5281/zenodo.8140487}.

\appendix

\section{Appendix A: Robustness of Timestepping Scheme}\label{sec:rung_ecc}

As described in section \ref{sec:methods}, {\sc ChaNGa}
  evolves the motions of the particles in the planetesimal disk using
  a multi-tiered timestepping scheme. Due to the extremely short
  dynamical timescale at the inner edge of the disk, the outer disk
  would require a prohibitive number of timesteps to reach the
  protoplanet phase using a fixed timestep scheme.  To circumvent this, particles are evolved in discrete power-of-two timestep groups. In the event that a collision occurs between two particles on different timesteps, a slight error is introduced to the energy and angular momentum of the merged particle. Due to the nonlinear nature of the runaway growth phase, this slight error tends to trigger more subsequent collisions at the timestep boundary in the disk, and causes protoplanets to preferentially form at the boundaries.

To circumvent this issue, we prohibit particles on different timesteps from merging until the runaway growth phase is well underway. For the fdHi simulation, multi-tiered mergers are not allowed during the first thousand steps. To verify that this technique does not alter the resulting protoplanet distribution in any meaningful way, we ran two test simulations of the inner part (1 to 4 days in orbital period) of the disk from the fdHi simulation. In the first case, the aforementioned timestepping scheme is used. In the second case, all particles are evolved on the timestep appropriate for the inner edge of the disk.

\begin{figure}
\begin{center}
    \includegraphics[width=0.5\textwidth]{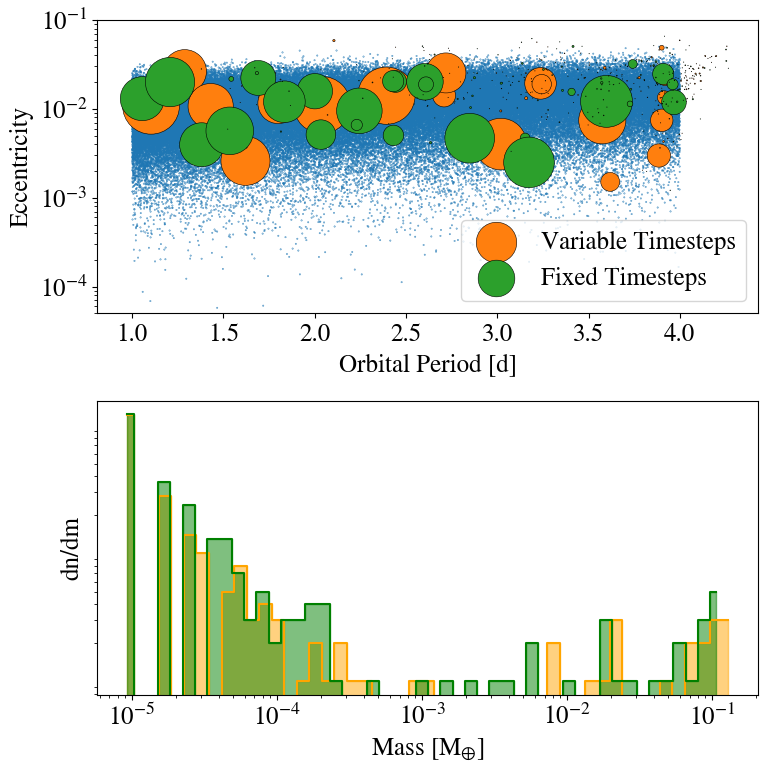}
    \caption{A comparison between the innermost region of the fdHi (orange) simulation, and a second version using a fixed timestep appropriate for the inner edge of the disk (green). In the top panel, the period-eccentricity state of the particles is shown, with marker sizes indicating relative mass. The blue points represent the initial state of the simulations. The bottom panel compares the final mass distributions of the bodies.\label{fig:rung_ecc}}
\end{center}
\end{figure}

In figure \ref{fig:rung_ecc}, we compare the final period-eccentricity state and final mass distributions to each other. There do not appear to be any differences between the two protoplanet distributions, particularly near the timestep boundary at 2 days. In addition, the masses of both the protoplanets and the remaining growing planetesimals are indistinguishable. In this case, a KS test of the two mass distributions yields a p-value of $\sim 0.34$. We therefore conclude that the timestepping scheme used in this work does not alter the growth of the protoplanets in any meaningful way.

\bibliography{references}

\begin{thebibliography}{}
\expandafter\ifx\csname natexlab\endcsname\relax\def\natexlab#1{#1}\fi
\providecommand{\url}[1]{\href{#1}{#1}}
\providecommand{\dodoi}[1]{doi:~\href{http://doi.org/#1}{\nolinkurl{#1}}}
\providecommand{\doeprint}[1]{\href{http://ascl.net/#1}{\nolinkurl{http://ascl.net/#1}}}
\providecommand{\doarXiv}[1]{\href{https://arxiv.org/abs/#1}{\nolinkurl{https://arxiv.org/abs/#1}}}

\bibitem[{{Adachi} {et~al.}(1976){Adachi}, {Hayashi}, \& {Nakazawa}}]{adachi76}
{Adachi}, I., {Hayashi}, C., \& {Nakazawa}, K. 1976, Progress of Theoretical
  Physics, 56, 1756, \dodoi{10.1143/PTP.56.1756}

\bibitem[{{Agol} {et~al.}(2021){Agol}, {Dorn}, {Grimm}, {Turbet}, {Ducrot},
  {Delrez}, {Gillon}, {Demory}, {Burdanov}, {Barkaoui}, {Benkhaldoun},
  {Bolmont}, {Burgasser}, {Carey}, {de Wit}, {Fabrycky}, {Foreman-Mackey},
  {Haldemann}, {Hernandez}, {Ingalls}, {Jehin}, {Langford}, {Leconte},
  {Lederer}, {Luger}, {Malhotra}, {Meadows}, {Morris}, {Pozuelos}, {Queloz},
  {Raymond}, {Selsis}, {Sestovic}, {Triaud}, \& {Van Grootel}}]{agol21}
{Agol}, E., {Dorn}, C., {Grimm}, S.~L., {et~al.} 2021, \psj, 2, 1,
  \dodoi{10.3847/PSJ/abd022}

\bibitem[{{Andrews} \& {Williams}(2005)}]{andrews05}
{Andrews}, S.~M., \& {Williams}, J.~P. 2005, \apj, 631, 1134,
  \dodoi{10.1086/432712}

\bibitem[{{Andrews} {et~al.}(2009){Andrews}, {Wilner}, {Hughes}, {Qi}, \&
  {Dullemond}}]{andrews09}
{Andrews}, S.~M., {Wilner}, D.~J., {Hughes}, A.~M., {Qi}, C., \& {Dullemond},
  C.~P. 2009, \apj, 700, 1502, \dodoi{10.1088/0004-637X/700/2/1502}

\bibitem[{{Andrews} {et~al.}(2010){Andrews}, {Wilner}, {Hughes}, {Qi}, \&
  {Dullemond}}]{andrews10}
---. 2010, \apj, 723, 1241, \dodoi{10.1088/0004-637X/723/2/1241}

\bibitem[{{Armitage}(2020)}]{armitage20}
{Armitage}, P.~J. 2020, {Astrophysics of planet formation, Second Edition}

\bibitem[{{Astropy Collaboration} {et~al.}(2013){Astropy Collaboration},
  {Robitaille}, {Tollerud}, {Greenfield}, {Droettboom}, {Bray}, {Aldcroft},
  {Davis}, {Ginsburg}, {Price-Whelan}, {Kerzendorf}, {Conley}, {Crighton},
  {Barbary}, {Muna}, {Ferguson}, {Grollier}, {Parikh}, {Nair}, {Unther},
  {Deil}, {Woillez}, {Conseil}, {Kramer}, {Turner}, {Singer}, {Fox}, {Weaver},
  {Zabalza}, {Edwards}, {Azalee Bostroem}, {Burke}, {Casey}, {Crawford},
  {Dencheva}, {Ely}, {Jenness}, {Labrie}, {Lim}, {Pierfederici}, {Pontzen},
  {Ptak}, {Refsdal}, {Servillat}, \& {Streicher}}]{astropy13}
{Astropy Collaboration}, {Robitaille}, T.~P., {Tollerud}, E.~J., {et~al.} 2013,
  \aap, 558, A33, \dodoi{10.1051/0004-6361/201322068}

\bibitem[{{Bai} \& {Stone}(2010)}]{bai10}
{Bai}, X.-N., \& {Stone}, J.~M. 2010, \apj, 722, 1437,
  \dodoi{10.1088/0004-637X/722/2/1437}

\bibitem[{{Baraffe} {et~al.}(2015){Baraffe}, {Homeier}, {Allard}, \&
  {Chabrier}}]{baraffe15}
{Baraffe}, I., {Homeier}, D., {Allard}, F., \& {Chabrier}, G. 2015, \aap, 577,
  A42, \dodoi{10.1051/0004-6361/201425481}

\bibitem[{{Barnes} {et~al.}(2009){Barnes}, {Quinn}, {Lissauer}, \&
  {Richardson}}]{barnes09}
{Barnes}, R., {Quinn}, T.~R., {Lissauer}, J.~J., \& {Richardson}, D.~C. 2009,
  \icarus, 203, 626, \dodoi{10.1016/j.icarus.2009.03.042}

\bibitem[{{Beitz} {et~al.}(2011){Beitz}, {G{\"u}ttler}, {Blum}, {Meisner},
  {Teiser}, \& {Wurm}}]{beitz11}
{Beitz}, E., {G{\"u}ttler}, C., {Blum}, J., {et~al.} 2011, \apj, 736, 34,
  \dodoi{10.1088/0004-637X/736/1/34}

\bibitem[{{Birnstiel} {et~al.}(2016){Birnstiel}, {Fang}, \&
  {Johansen}}]{birnstiel16}
{Birnstiel}, T., {Fang}, M., \& {Johansen}, A. 2016, \ssr, 205, 41,
  \dodoi{10.1007/s11214-016-0256-1}

\bibitem[{{Bitsch} {et~al.}(2015){Bitsch}, {Lambrechts}, \&
  {Johansen}}]{bitsch15}
{Bitsch}, B., {Lambrechts}, M., \& {Johansen}, A. 2015, \aap, 582, A112,
  \dodoi{10.1051/0004-6361/201526463}

\bibitem[{{Blum} \& {M{\"u}nch}(1993)}]{blum93}
{Blum}, J., \& {M{\"u}nch}, M. 1993, \icarus, 106, 151,
  \dodoi{10.1006/icar.1993.1163}

\bibitem[{{Bottke} {et~al.}(2005){Bottke}, {Durda}, {Nesvorn{\'y}}, {Jedicke},
  {Morbidelli}, {Vokrouhlick{\'y}}, \& {Levison}}]{bottke05}
{Bottke}, W.~F., {Durda}, D.~D., {Nesvorn{\'y}}, D., {et~al.} 2005, \icarus,
  179, 63, \dodoi{10.1016/j.icarus.2005.05.017}

\bibitem[{{Brauer} {et~al.}(2008){Brauer}, {Henning}, \&
  {Dullemond}}]{brauer08b}
{Brauer}, F., {Henning}, T., \& {Dullemond}, C.~P. 2008, \aap, 487, L1,
  \dodoi{10.1051/0004-6361:200809780}

\bibitem[{{Chambers} \& {Wetherill}(1998)}]{chambers98}
{Chambers}, J.~E., \& {Wetherill}, G.~W. 1998, \icarus, 136, 304,
  \dodoi{10.1006/icar.1998.6007}

\bibitem[{{Chambers} \& {Wetherill}(2001)}]{chambers01}
---. 2001, \maps, 36, 381, \dodoi{10.1111/j.1945-5100.2001.tb01881.x}

\bibitem[{{Chiang} \& {Laughlin}(2013)}]{chiang13}
{Chiang}, E., \& {Laughlin}, G. 2013, \mnras, 431, 3444,
  \dodoi{10.1093/mnras/stt424}

\bibitem[{{Childs} \& {Steffen}(2022)}]{childs22}
{Childs}, A.~C., \& {Steffen}, J.~H. 2022, \mnras, 511, 1848,
  \dodoi{10.1093/mnras/stac158}

\bibitem[{{Clement} {et~al.}(2020){Clement}, {Kaib}, \& {Chambers}}]{clement20}
{Clement}, M.~S., {Kaib}, N.~A., \& {Chambers}, J.~E. 2020, PSJ, 1, 18,
  \dodoi{10.3847/PSJ/ab91aa}

\bibitem[{{Colwell}(2003)}]{colwell03}
{Colwell}, J.~E. 2003, \icarus, 164, 188, \dodoi{10.1016/S0019-1035(03)00083-6}

\bibitem[{{Drk{a}{\.z}kowska} \& {Dullemond}(2018)}]{drkazowska18}
{Drk{a}{\.z}kowska}, J., \& {Dullemond}, C.~P. 2018, \aap, 614, A62,
  \dodoi{10.1051/0004-6361/201732221}

\bibitem[{{Drk{a}{\.z}kowska} {et~al.}(2013){Drk{a}{\.z}kowska}, {Windmark}, \&
  {Dullemond}}]{drkazowska13}
{Drk{a}{\.z}kowska}, J., {Windmark}, F., \& {Dullemond}, C.~P. 2013, \aap, 556,
  A37, \dodoi{10.1051/0004-6361/201321566}

\bibitem[{{Duncan} {et~al.}(1989){Duncan}, {Quinn}, \& {Tremaine}}]{duncan89}
{Duncan}, M., {Quinn}, T., \& {Tremaine}, S. 1989, \icarus, 82, 402,
  \dodoi{10.1016/0019-1035(89)90047-X}

\bibitem[{{Emsenhuber} {et~al.}(2021{\natexlab{a}}){Emsenhuber}, {Mordasini},
  {Burn}, {Alibert}, {Benz}, \& {Asphaug}}]{emsenhuber21a}
{Emsenhuber}, A., {Mordasini}, C., {Burn}, R., {et~al.} 2021{\natexlab{a}},
  \aap, 656, A69, \dodoi{10.1051/0004-6361/202038553}

\bibitem[{{Emsenhuber} {et~al.}(2021{\natexlab{b}}){Emsenhuber}, {Mordasini},
  {Burn}, {Alibert}, {Benz}, \& {Asphaug}}]{emsenhuber21b}
---. 2021{\natexlab{b}}, \aap, 656, A70, \dodoi{10.1051/0004-6361/202038863}

\bibitem[{{Fabrycky} {et~al.}(2014){Fabrycky}, {Lissauer}, {Ragozzine}, {Rowe},
  {Steffen}, {Agol}, {Barclay}, {Batalha}, {Borucki}, \& {Ciardi}}]{fabrycky14}
{Fabrycky}, D.~C., {Lissauer}, J.~J., {Ragozzine}, D., {et~al.} 2014, \apj,
  790, 146, \dodoi{10.1088/0004-637X/790/2/146}

\bibitem[{{Gillon} {et~al.}(2016){Gillon}, {Jehin}, {Lederer}, {Delrez}, {de
  Wit}, {Burdanov}, {Van Grootel}, {Burgasser}, {Triaud}, {Opitom}, {Demory},
  {Sahu}, {Bardalez Gagliuffi}, {Magain}, \& {Queloz}}]{gillon16}
{Gillon}, M., {Jehin}, E., {Lederer}, S.~M., {et~al.} 2016, \nat, 533, 221,
  \dodoi{10.1038/nature17448}

\bibitem[{{Gillon} {et~al.}(2017){Gillon}, {Triaud}, {Demory}, {Jehin}, {Agol},
  {Deck}, {Lederer}, {de Wit}, {Burdanov}, {Ingalls}, {Bolmont}, {Leconte},
  {Raymond}, {Selsis}, {Turbet}, {Barkaoui}, {Burgasser}, {Burleigh}, {Carey},
  {Chaushev}, {Copperwheat}, {Delrez}, {Fernandes}, {Holdsworth}, {Kotze}, {Van
  Grootel}, {Almleaky}, {Benkhaldoun}, {Magain}, \& {Queloz}}]{gillon17}
{Gillon}, M., {Triaud}, A. H.~M.~J., {Demory}, B.-O., {et~al.} 2017, \nat, 542,
  456, \dodoi{10.1038/nature21360}

\bibitem[{{Hansen} \& {Murray}(2012)}]{hansen12}
{Hansen}, B. M.~S., \& {Murray}, N. 2012, \apj, 751, 158,
  \dodoi{10.1088/0004-637X/751/2/158}

\bibitem[{{Hayashi}(1981)}]{hayashi81}
{Hayashi}, C. 1981, Progress of Theoretical Physics Supplement, 70, 35,
  \dodoi{10.1143/PTPS.70.35}

\bibitem[{{Hunter}(2007)}]{matplotlib07}
{Hunter}, J.~D. 2007, Computing in Science and Engineering, 9, 90,
  \dodoi{10.1109/MCSE.2007.55}

\bibitem[{{Ida}(1990)}]{ida90}
{Ida}, S. 1990, \icarus, 88, 129, \dodoi{10.1016/0019-1035(90)90182-9}

\bibitem[{{Ida} {et~al.}(1997){Ida}, {Canup}, \& {Stewart}}]{ida97}
{Ida}, S., {Canup}, R.~M., \& {Stewart}, G.~R. 1997, \nat, 389, 353,
  \dodoi{10.1038/38669}

\bibitem[{{Ida} {et~al.}(1993){Ida}, {Kokubo}, \& {Makino}}]{ida93a}
{Ida}, S., {Kokubo}, E., \& {Makino}, J. 1993, \mnras, 263, 875,
  \dodoi{10.1093/mnras/263.4.875}

\bibitem[{{Ida} \& {Makino}(1993)}]{ida93}
{Ida}, S., \& {Makino}, J. 1993, \icarus, 106, 210,
  \dodoi{10.1006/icar.1993.1167}

\bibitem[{{Ida} {et~al.}(2020){Ida}, {Ueta}, {Sasaki}, \& {Ishizawa}}]{ida20}
{Ida}, S., {Ueta}, S., {Sasaki}, T., \& {Ishizawa}, Y. 2020, Nature Astronomy,
  4, 880, \dodoi{10.1038/s41550-020-1049-8}

\bibitem[{{Izidoro} {et~al.}(2021){Izidoro}, {Bitsch}, {Raymond}, {Johansen},
  {Morbidelli}, {Lambrechts}, \& {Jacobson}}]{izidoro21}
{Izidoro}, A., {Bitsch}, B., {Raymond}, S.~N., {et~al.} 2021, \aap, 650, A152,
  \dodoi{10.1051/0004-6361/201935336}

\bibitem[{{Izidoro} {et~al.}(2017){Izidoro}, {Ogihara}, {Raymond},
  {Morbidelli}, {Pierens}, {Bitsch}, {Cossou}, \& {Hersant}}]{izidoro17}
{Izidoro}, A., {Ogihara}, M., {Raymond}, S.~N., {et~al.} 2017, \mnras, 470,
  1750, \dodoi{10.1093/mnras/stx1232}

\bibitem[{{Jetley} {et~al.}(2008){Jetley}, {Gioachin}, {Mendes}, \&
  {Quinn}}]{jetley08}
{Jetley}, P., {Gioachin}, F., {Mendes}, C.~{Kale}, L., \& {Quinn}, T. 2008,
  Proceedings of IEEE International Parallel and Distributed Processing
  Symposium

\bibitem[{{Johansen} {et~al.}(2014){Johansen}, {Blum}, {Tanaka}, {Ormel},
  {Bizzarro}, \& {Rickman}}]{johansen14}
{Johansen}, A., {Blum}, J., {Tanaka}, H., {et~al.} 2014, in Protostars and
  Planets VI, ed. H.~{Beuther}, R.~S. {Klessen}, C.~P. {Dullemond}, \&
  T.~{Henning}, 547

\bibitem[{{Johansen} {et~al.}(2007){Johansen}, {Oishi}, {Mac Low}, {Klahr},
  {Henning}, \& {Youdin}}]{johansen07}
{Johansen}, A., {Oishi}, J.~S., {Mac Low}, M.-M., {et~al.} 2007, \nat, 448,
  1022, \dodoi{10.1038/nature06086}

\bibitem[{{Johansen} {et~al.}(2009){Johansen}, {Youdin}, \&
  {Klahr}}]{johansen2009b}
{Johansen}, A., {Youdin}, A., \& {Klahr}, H. 2009, \apj, 697, 1269,
  \dodoi{10.1088/0004-637X/697/2/1269}

\bibitem[{{Kauffmann} \& {White}(1993)}]{kauffmann93}
{Kauffmann}, G., \& {White}, S.~D.~M. 1993, \mnras, 261, 921,
  \dodoi{10.1093/mnras/261.4.921}

\bibitem[{{Klahr} \& {Bodenheimer}(2003)}]{klahr03}
{Klahr}, H.~H., \& {Bodenheimer}, P. 2003, \apj, 582, 869,
  \dodoi{10.1086/344743}

\bibitem[{{Kokubo} \& {Ida}(1995)}]{kokubo95}
{Kokubo}, E., \& {Ida}, S. 1995, \icarus, 114, 247,
  \dodoi{10.1006/icar.1995.1059}

\bibitem[{{Kokubo} \& {Ida}(1996)}]{kokubo96}
---. 1996, \icarus, 123, 180, \dodoi{10.1006/icar.1996.0148}

\bibitem[{{Kokubo} \& {Ida}(1998)}]{kokubo98}
---. 1998, \icarus, 131, 171, \dodoi{10.1006/icar.1997.5840}

\bibitem[{{Kokubo} \& {Ida}(2000)}]{kokubo00}
---. 2000, \icarus, 143, 15, \dodoi{10.1006/icar.1999.6237}

\bibitem[{{Kokubo} \& {Ida}(2002)}]{kokubo02}
---. 2002, \apj, 581, 666, \dodoi{10.1086/344105}

\bibitem[{{Kokubo} {et~al.}(2000){Kokubo}, {Ida}, \& {Makino}}]{kokubo00b}
{Kokubo}, E., {Ida}, S., \& {Makino}, J. 2000, \icarus, 148, 419,
  \dodoi{10.1006/icar.2000.6496}

\bibitem[{{Lambrechts} \& {Johansen}(2012)}]{lambrechts12}
{Lambrechts}, M., \& {Johansen}, A. 2012, \aap, 544, A32,
  \dodoi{10.1051/0004-6361/201219127}

\bibitem[{{Lambrechts} \& {Johansen}(2014)}]{lambrechts14}
---. 2014, \aap, 572, A107, \dodoi{10.1051/0004-6361/201424343}

\bibitem[{{Latham} {et~al.}(2011){Latham}, {Rowe}, {Quinn}, {Batalha},
  {Borucki}, {Brown}, {Bryson}, {Buchhave}, {Caldwell}, {Carter},
  {Christiansen}, {Ciardi}, {Cochran}, {Dunham}, {Fabrycky}, {Ford}, {Gautier},
  {Gilliland}, {Holman}, {Howell}, {Ibrahim}, {Isaacson}, {Jenkins}, {Koch},
  {Lissauer}, {Marcy}, {Quintana}, {Ragozzine}, {Sasselov}, {Shporer},
  {Steffen}, {Welsh}, \& {Wohler}}]{lantham11}
{Latham}, D.~W., {Rowe}, J.~F., {Quinn}, S.~N., {et~al.} 2011, \apjl, 732, L24,
  \dodoi{10.1088/2041-8205/732/2/L24}

\bibitem[{{Leinhardt} {et~al.}(2015){Leinhardt}, {Dobinson}, {Carter}, \&
  {Lines}}]{leinhardt15}
{Leinhardt}, Z.~M., {Dobinson}, J., {Carter}, P.~J., \& {Lines}, S. 2015, \apj,
  806, 23, \dodoi{10.1088/0004-637X/806/1/23}

\bibitem[{{Leinhardt} \& {Richardson}(2005)}]{leinhardt05}
{Leinhardt}, Z.~M., \& {Richardson}, D.~C. 2005, \apj, 625, 427,
  \dodoi{10.1086/429402}

\bibitem[{{Leinhardt} \& {Stewart}(2012)}]{leinhardt12}
{Leinhardt}, Z.~M., \& {Stewart}, S.~T. 2012, \apj, 745, 79,
  \dodoi{10.1088/0004-637X/745/1/79}

\bibitem[{{Levison} {et~al.}(2011){Levison}, {Morbidelli}, {Tsiganis},
  {Nesvorn{\'y}}, \& {Gomes}}]{levison11}
{Levison}, H.~F., {Morbidelli}, A., {Tsiganis}, K., {Nesvorn{\'y}}, D., \&
  {Gomes}, R. 2011, \aj, 142, 152, \dodoi{10.1088/0004-6256/142/5/152}

\bibitem[{{Levison} {et~al.}(2008){Levison}, {Morbidelli}, {Van Laerhoven},
  {Gomes}, \& {Tsiganis}}]{levison08}
{Levison}, H.~F., {Morbidelli}, A., {Van Laerhoven}, C., {Gomes}, R., \&
  {Tsiganis}, K. 2008, \icarus, 196, 258, \dodoi{10.1016/j.icarus.2007.11.035}

\bibitem[{{L'Huillier} {et~al.}(2012){L'Huillier}, {Combes}, \&
  {Semelin}}]{lhullier12}
{L'Huillier}, B., {Combes}, F., \& {Semelin}, B. 2012, \aap, 544, A68,
  \dodoi{10.1051/0004-6361/201117924}

\bibitem[{{Lissauer}(1987)}]{lissauer87}
{Lissauer}, J.~J. 1987, \icarus, 69, 249, \dodoi{10.1016/0019-1035(87)90104-7}

\bibitem[{{Lissauer} \& {Stewart}(1993)}]{lissauer93}
{Lissauer}, J.~J., \& {Stewart}, G.~R. 1993, in Protostars and Planets III, ed.
  E.~H. {Levy} \& J.~I. {Lunine}, 1061

\bibitem[{{Lissauer} {et~al.}(2011){Lissauer}, {Ragozzine}, {Fabrycky},
  {Steffen}, {Ford}, {Jenkins}, {Shporer}, {Holman}, {Rowe}, {Quintana},
  {Batalha}, {Borucki}, {Bryson}, {Caldwell}, {Carter}, {Ciardi}, {Dunham},
  {Fortney}, {Gautier}, {Howell}, {Koch}, {Latham}, {Marcy}, {Morehead}, \&
  {Sasselov}}]{lissauer11}
{Lissauer}, J.~J., {Ragozzine}, D., {Fabrycky}, D.~C., {et~al.} 2011, \apjs,
  197, 8, \dodoi{10.1088/0067-0049/197/1/8}

\bibitem[{{Lithwick}(2014)}]{lithwick14}
{Lithwick}, Y. 2014, \apj, 780, 22, \dodoi{10.1088/0004-637X/780/1/22}

\bibitem[{{Lyra} {et~al.}(2008){Lyra}, {Johansen}, {Klahr}, \&
  {Piskunov}}]{lyra08}
{Lyra}, W., {Johansen}, A., {Klahr}, H., \& {Piskunov}, N. 2008, \aap, 491,
  L41, \dodoi{10.1051/0004-6361:200810626}

\bibitem[{{Mamajek}(2009)}]{mamajek09}
{Mamajek}, E.~E. 2009, in American Institute of Physics Conference Series, Vol.
  1158, Exoplanets and Disks: Their Formation and Diversity, ed. T.~{Usuda},
  M.~{Tamura}, \& M.~{Ishii}, 3--10

\bibitem[{{Menon} {et~al.}(2015){Menon}, {Wesolowski}, \& {Zheng}}]{menon15}
{Menon}, H., {Wesolowski}, L., \& {Zheng}, G. e.~a. 2015, Computational
  Astrophysics and Cosmology, 2, 1

\bibitem[{{Morbidelli} {et~al.}(2009){Morbidelli}, {Bottke}, {Nesvorn{\'y}}, \&
  {Levison}}]{morbidelli09}
{Morbidelli}, A., {Bottke}, W.~F., {Nesvorn{\'y}}, D., \& {Levison}, H.~F.
  2009, \icarus, 204, 558, \dodoi{10.1016/j.icarus.2009.07.011}

\bibitem[{{Morishima} {et~al.}(2010){Morishima}, {Stadel}, \&
  {Moore}}]{morishima10}
{Morishima}, R., {Stadel}, J., \& {Moore}, B. 2010, \icarus, 207, 517,
  \dodoi{10.1016/j.icarus.2009.11.038}

\bibitem[{{Mulders} {et~al.}(2018){Mulders}, {Pascucci}, {Apai}, \&
  {Ciesla}}]{mulders18}
{Mulders}, G.~D., {Pascucci}, I., {Apai}, D., \& {Ciesla}, F.~J. 2018, \aj,
  156, 24, \dodoi{10.3847/1538-3881/aac5ea}

\bibitem[{{Mundy} {et~al.}(2000){Mundy}, {Looney}, \& {Welch}}]{mundy00}
{Mundy}, L.~G., {Looney}, L.~W., \& {Welch}, W.~J. 2000, in Protostars and
  Planets IV, ed. V.~{Mannings}, A.~P. {Boss}, \& S.~S. {Russell}, 355

\bibitem[{{Murali} {et~al.}(2002){Murali}, {Katz}, {Hernquist}, {Weinberg}, \&
  {Dav{\'e}}}]{murali02}
{Murali}, C., {Katz}, N., {Hernquist}, L., {Weinberg}, D.~H., \& {Dav{\'e}}, R.
  2002, \apj, 571, 1, \dodoi{10.1086/339876}

\bibitem[{{Nakagawa} {et~al.}(1986){Nakagawa}, {Sekiya}, \&
  {Hayashi}}]{nakagawa86}
{Nakagawa}, Y., {Sekiya}, M., \& {Hayashi}, C. 1986, \icarus, 67, 375,
  \dodoi{10.1016/0019-1035(86)90121-1}

\bibitem[{{Nakazawa} \& {Ida}(1988)}]{naka88}
{Nakazawa}, K., \& {Ida}, S. 1988, Progress of Theoretical Physics Supplement,
  96, 167, \dodoi{10.1143/PTPS.96.167}

\bibitem[{{Pontzen} {et~al.}(2013){Pontzen}, {Ro{\v{s}}kar}, {Stinson}, \&
  {Woods}}]{pynbody13}
{Pontzen}, A., {Ro{\v{s}}kar}, R., {Stinson}, G., \& {Woods}, R. 2013,
  {pynbody: N-Body/SPH analysis for python}.
\newblock \doeprint{1305.002}

\bibitem[{{Raymond} {et~al.}(2009){Raymond}, {O'Brien}, {Morbidelli}, \&
  {Kaib}}]{raymond09}
{Raymond}, S.~N., {O'Brien}, D.~P., {Morbidelli}, A., \& {Kaib}, N.~A. 2009,
  \icarus, 203, 644, \dodoi{10.1016/j.icarus.2009.05.016}

\bibitem[{{Raymond} {et~al.}(2005){Raymond}, {Quinn}, \& {Lunine}}]{raymond05}
{Raymond}, S.~N., {Quinn}, T., \& {Lunine}, J.~I. 2005, \apj, 632, 670,
  \dodoi{10.1086/433179}

\bibitem[{{Raymond} {et~al.}(2006){Raymond}, {Quinn}, \& {Lunine}}]{raymond06}
---. 2006, \icarus, 183, 265, \dodoi{10.1016/j.icarus.2006.03.011}

\bibitem[{{Raymond} {et~al.}(2007){Raymond}, {Scalo}, \& {Meadows}}]{raymond07}
{Raymond}, S.~N., {Scalo}, J., \& {Meadows}, V.~S. 2007, \apj, 669, 606,
  \dodoi{10.1086/521587}

\bibitem[{{Richardson}(1994)}]{richardson94}
{Richardson}, D.~C. 1994, \mnras, 269, 493, \dodoi{10.1093/mnras/269.2.493}

\bibitem[{{Richardson} {et~al.}(2000){Richardson}, {Quinn}, {Stadel}, \&
  {Lake}}]{richardson00}
{Richardson}, D.~C., {Quinn}, T., {Stadel}, J., \& {Lake}, G. 2000, \icarus,
  143, 45, \dodoi{10.1006/icar.1999.6243}

\bibitem[{{Rowe} {et~al.}(2014){Rowe}, {Bryson}, {Marcy}, {Lissauer},
  {Jontof-Hutter}, {Mullally}, {Gilliland}, {Issacson}, {Ford}, {Howell},
  {Borucki}, {Haas}, {Huber}, {Steffen}, {Thompson}, {Quintana}, {Barclay},
  {Still}, {Fortney}, {Gautier}, {Hunter}, {Caldwell}, {Ciardi}, {Devore},
  {Cochran}, {Jenkins}, {Agol}, {Carter}, \& {Geary}}]{rowe14}
{Rowe}, J.~F., {Bryson}, S.~T., {Marcy}, G.~W., {et~al.} 2014, \apj, 784, 45,
  \dodoi{10.1088/0004-637X/784/1/45}

\bibitem[{{Safronov}(1969)}]{safronov69}
{Safronov}, V.~S. 1969, {Evoliutsiia doplanetnogo oblaka.}

\bibitem[{{Sheppard} \& {Trujillo}(2010)}]{sheppard10}
{Sheppard}, S.~S., \& {Trujillo}, C.~A. 2010, \apjl, 723, L233,
  \dodoi{10.1088/2041-8205/723/2/L233}

\bibitem[{{Shibaike} \& {Alibert}(2020)}]{shibaike20}
{Shibaike}, Y., \& {Alibert}, Y. 2020, \aap, 644, A81,
  \dodoi{10.1051/0004-6361/202039086}

\bibitem[{{Simon} {et~al.}(2012){Simon}, {Beckwith}, \& {Armitage}}]{simon12}
{Simon}, J.~B., {Beckwith}, K., \& {Armitage}, P.~J. 2012, \mnras, 422, 2685,
  \dodoi{10.1111/j.1365-2966.2012.20835.x}

\bibitem[{{Stewart} \& {Ida}(2000)}]{stewart00}
{Stewart}, G.~R., \& {Ida}, S. 2000, \icarus, 143, 28,
  \dodoi{10.1006/icar.1999.6242}

\bibitem[{{van der Walt} {et~al.}(2011){van der Walt}, {Colbert}, \&
  {Varoquaux}}]{numpy11}
{van der Walt}, S., {Colbert}, S.~C., \& {Varoquaux}, G. 2011, Computing in
  Science and Engineering, 13, 22, \dodoi{10.1109/MCSE.2011.37}

\bibitem[{{Wallace} \& {Quinn}(2019)}]{wallace19}
{Wallace}, S.~C., \& {Quinn}, T.~R. 2019, \mnras, 489, 2159,
  \dodoi{10.1093/mnras/stz2284}

\bibitem[{{Weidenschilling}(1977)}]{weidenschilling77}
{Weidenschilling}, S.~J. 1977, \mnras, 180, 57, \dodoi{10.1093/mnras/180.2.57}

\bibitem[{{Weidenschilling}(1989)}]{weidenschilling89}
---. 1989, \icarus, 80, 179, \dodoi{10.1016/0019-1035(89)90166-8}

\bibitem[{{W}es {M}c{K}inney(2010)}]{pandas10}
{W}es {M}c{K}inney. 2010, in {P}roceedings of the 9th {P}ython in {S}cience
  {C}onference, ed. {S}t\'efan van~der {W}alt \& {J}arrod {M}illman, 56 -- 61

\bibitem[{{Wetherill} \& {Stewart}(1989)}]{wetherill89}
{Wetherill}, G.~W., \& {Stewart}, G.~R. 1989, \icarus, 77, 330,
  \dodoi{10.1016/0019-1035(89)90093-6}

\bibitem[{{Wetherill} \& {Stewart}(1993)}]{wetherill93}
---. 1993, \icarus, 106, 190, \dodoi{10.1006/icar.1993.1166}

\end{thebibliography}

\clearpage

\end{document}